\documentclass[acmsmall]{acmart}

\usepackage{amsmath}
\usepackage[makeroom]{cancel}   
\usepackage{tikz}               
\usepackage[inline]{enumitem}   
\usepackage{makecell}  
\usepackage{multirow}  
\usepackage{diagbox}   
\usepackage{subcaption}  
\usepackage[frozencache,cachedir=minted-cache]{minted}    
\usepackage[ruled,linesnumbered,noend]{algorithm2e}
\usepackage{xparse}  
\usepackage[dvipsnames]{xcolor}  
\usepackage{pifont}  
\usepackage[normalem]{ulem}  
\usepackage{adjustbox}  
\usepackage{wrapfig}    

\newcommand{\NAME}{Neptune}

\setminted{breaklines=true, breakanywhere=true, fontsize=\footnotesize}

\setlist[itemize]{leftmargin=1em, topsep=1pt, partopsep=1pt}
\setlist[enumerate]{leftmargin=1em, topsep=1pt, partopsep=1pt}
\setlist[description]{leftmargin=1em, topsep=1pt, partopsep=1pt}
\makeatletter
\patchcmd{\@algocf@start}{\addtolength{\hsize}{-\algomargin}\addtolength{\hsize}{-1.5em}}
{\addtolength{\hsize}{-\algomargin}\addtolength{\hsize}{-0.4em}}{}{}
\makeatother

\ifdefined\SepAppendix
\newcommand{\appref}[1]{\href{https://github.com/uiuc-arc/neptune/tree/main/_paper/neptune-appendix.pdf}{\ref*{#1}}}
\newcommand{\appcite}{\cite{neptune-appendix}}
\else
\newcommand{\appref}[1]{\ref{#1}}
\newcommand{\appcite}{}
\fi

\setcopyright{cc}
\setcctype{by}
\acmDOI{10.1145/3808298}
\acmYear{2026}
\acmJournal{PACMPL}
\acmVolume{10}
\acmNumber{PLDI}
\acmArticle{220}
\acmMonth{6}
\acmSubmissionID{pldi26main-p347-p}
\received{2025-11-14}
\received[accepted]{2026-04-03}

\title{\texorpdfstring{Neptune: Advanced ML Operator Fusion \\for Locality and Parallelism on GPUs
  }{Neptune: Advanced ML Operator Fusion for Locality and Parallelism on GPUs
}}
\renewcommand{\shorttitle}{Neptune: Advanced ML Operator Fusion for Locality and
Parallelism on GPUs}

\author{Yifan Zhao}
\orcid{0009-0007-7080-891X}
\affiliation{%
  \institution{University of Illinois Urbana-Champaign}
  \country{USA}
}
\email{yifanz16@illinois.edu}

\author{Egan Johnson}
\orcid{0000-0002-2989-2212}
\affiliation{%
  \institution{University of Illinois Urbana-Champaign}
  \country{USA}
}
\email{egancj2@illinois.edu}

\author{Prasanth Chatarasi}
\orcid{0000-0002-0974-4001}
\affiliation{%
  \institution{IBM Research}
  \country{USA}
}
\email{prasanth@ibm.com}

\author{Vikram S. Adve}
\orcid{0000-0002-0760-9690}
\affiliation{%
  \institution{University of Illinois Urbana-Champaign}
  \country{USA}
}
\email{vadve@illinois.edu}

\author{Sasa Misailovic}
\orcid{0000-0001-7319-8845}
\affiliation{%
  \institution{University of Illinois Urbana-Champaign}
  \country{USA}
}
\email{misailo@illinois.edu}

\newcommand{\numberthis}{\addtocounter{equation}{1}\tag{\theequation}}

\def\ssum{\texttt{s\_sum}}
\def\smax{\texttt{s\_max}}
\def\sexp{\texttt{s\_exp}}
\def\loopj{\texttt{loop\_j}}
\def\inp{\texttt{inp}}
\def\xmax{\texttt{xmax}}
\def\xmaxi{\texttt{\xmax[i]}}
\def\xexp{\texttt{xexp}}
\def\xsum{\texttt{xsum}}
\def\xsumi{\texttt{\xsum[i]}}
\def\xmaxprev{\texttt{xmax\_0}}
\def\xmaxprevi{\texttt{\xmaxprev[i]}}
\def\xmaxcurr{\texttt{xmax\_1}}
\def\xmaxcurri{\texttt{\xmaxcurr[i]}}
\newcommand{\tagged}[2]{\ensuremath{#1^{\langle#2\rangle}}}

\def\bi{\boldsymbol{i}}
\def\bj{\boldsymbol{j}}
\def\bn{\boldsymbol{n}}
\def\bm{\boldsymbol{m}}
\def\accf{\phi}
\newcommand{\access}[3]{#1[\accf_{#2}(#3)]}

\def\fold{\mathcal{R}}
\newcommand{\inext}[1]{\operatorname{next}(#1)}
\newcommand{\iprev}[1]{\operatorname{prev}(#1)}
\def\nextj{\inext{\bj}}
\def\prevj{\iprev{\bj}}

\newcommand{\dom}[1]{\mathcal{D}^{#1}}
\newcommand{\circled}[1]{\ \tikz[baseline=(char.base)]{\node[shape=circle,draw,inner
sep=0.5pt] (char) {\footnotesize $#1$};}\ }
\def\cf{\circled{f}}
\newcommand{\rnest}[5]{rnest(#1:#2, #3:#4)\left\{\ #5\ \right\}}

\newcommand{\original}[1]{\textcolor[HTML]{4E95D9}{#1}}
\newcommand{\incorrect}[1]{\textcolor[HTML]{B25628}{#1}}
\newcommand{\fixed}[1]{\textcolor[HTML]{278117}{#1}}

\begin{document}

\begin{abstract}
  Operator fusion has become a key optimization for deep learning,
which combines multiple deep learning operators to improve data reuse and reduce global
memory transfers.
However, existing tensor compilers struggle to fuse complex reduction computations
involving loop-carried dependencies, \mbox{such as attention mechanisms.}

This paper introduces Neptune, a tensor compiler for advanced operator fusion for
sequences of reduction operators.
Neptune presents a new approach for advanced operator fusion, which intentionally
breaks some existing dependencies and compensates by constructing algebraic correction expressions
that allow the kernel to produce the correct result. Applying Neptune’s advanced operator
fusion to a plain attention operator generates operators
equivalent to FlashAttention and FlashDecoding.

On ten attention-based benchmarks, Neptune, starting from a plain attention code
and a high-level scheduling template, outperforms existing compilers
like Triton, TVM, and FlexAttention, including Triton-based implementations of FlashAttention.
Across four different GPU architectures from NVIDIA and AMD,
Neptune-generated kernels have an average speedup of $1.35\times$ over the next best alternative,
with up to $2.65\times$ speedup on Nvidia GPUs and up to $3.32\times$ on AMD GPUs,
demonstrating its effectiveness for deep learning workloads.

\end{abstract}

\maketitle

\section{Introduction}
\label{sec:introduction}
{Finding high-performance implementations for modern deep learning models is essential
for low-latency and low-cost model deployment. There is a large gap between high-level
machine learning frameworks like Pytorch~\cite{pytorch} and TensorFlow~\cite{tensorflow},
which specify models in terms of mathematical tensor operators, and the high-performance
kernels near the hardware, which must account for performance characteristics like memory
hierarchies, complex loop tilings, and more.

Tensor kernel compilers are a promising way
to translate groups of tensor operators into efficient device kernels.
\emph{Tile-based compilers} and languages,
such as Triton~\cite{triton}, Pallas~\cite{pallas}, and TileLang~\cite{tilelang},
operate with a low level of abstraction over hardware vendor languages.
Programmers write tile-based programs where tiles (fixed-shape subtensors) are first-class objects.
Tile languages provide an SIMD interface,
and realize data-level parallelism on GPUs and accelerators.
However, tile languages' low level of abstraction means programming in tile languages
remains difficult,
requiring both significant hardware and  algorithmic optimization expertise.
\textit{Schedule-based compilers} such as Halide~\cite{halide} and TVM~\cite{tvm} provide
an alternative.
They combine a mathematical definition of the program,
with a recipe of transformation primitives (``schedule template'') to optimize the program.
As the schedule template calls built-in transformations, the compiler automatically
safeguards correctness and fidelity to the original program.

Presently, however, both tile and schedule compilers miss important transformations that
are critical for performance,
which forces programmers to escape the compiler pipeline and manually implement kernels
for important operators in tile frameworks or in vendor languages such as CUDA and HIP.
This approach may sometimes be effective, but requires significant expert effort and
sacrifices the portability, flexibility, and maintainability of high-level frameworks.

Attention operators are one example where the lack of key optimizations in a compiler
forces developers to provide handwritten kernels.
Attention is a sequence of four operations that benefit greatly from operator fusion:
matrix multiplication (matmul), element-wise division, softmax, and another matmul.
Tile languages are unable to provide this fusion automatically, as it is above the scope
of the tile optimizer and must be encoded by the developer in the input program.
Schedule languages are at the right abstraction level for operator fusion, but
existing compiler transformations are unable to fuse attention's operators due to complex
data dependencies surrounding multiple reduction loops.
As a result, compilers and developers resort to manually developed, specialized fusion solutions
such as FlashAttention~\cite{flash-attention,flash-attention-2} and
FlashDecoding~\cite{flash-decoding}, which struggle to generalize to diverse workloads of
variant operators.

We identify two missing pieces preventing current tensor compilers from capturing
optimizations equivalent to FlashAttention and FlashDecoding:
\begin{itemize}
  \item Standard loop fusion in today's tensor compilers misses many optimization opportunities
    because of complex, fusion-preventing dependencies.
    Programmers divert from the compiler pipeline to apply manual, ad-hoc operator fusion,
    and imitate the code generation capabilities of compilers with code templates like
    FlexAttention.
    This situation indicates that generalizing operator fusion is necessary and possible,
    but existing tensor compilers are not capable of it.
  \item Existing tensor optimization approaches do not suffice alone.
    Tile optimizers leave important high-level optimizations --- such as operator fusion ---
    to programmers.
    Schedule optimizers produce intractable search spaces,
    with long sequences of many fine-grained transformations,
    that overwhelm programmers and autotuners.
    Integration of schedule and tile optimization pipelines is desirable,
    but has been virtually non-existent.
\end{itemize}
\vspace{0.3em}
\noindent\textbf{Our Work:}
We present \NAME{}, a novel tensor kernel compiler that combines novel advanced operator
fusion with a schedule optimizer and tile optimizer to provide full optimization in one pipeline.
\NAME{} fuses multiple reduction operators with complex data dependencies
that are beyond the reach of existing popular tensor compilers
such as TVM~\cite{tvm}, Triton~\cite{triton}, and OpenXLA~\cite{xla}.
Applying \NAME{}'s advanced operator fusion to a plain attention operator generates
operators equivalent to FlashAttention and FlashDecoding.
\NAME{} makes two key technical innovations:

\begin{itemize}
  \item We present a novel paradigm for advanced operator fusion with sequences of
    reduction operators.
    This paradigm solves the issue of fusion-preventing data dependencies by
  \begin{enumerate*}[label=\arabic*)]
  \item intentionally breaking some of those dependencies (``naive fusion'') and
  \item constructing algebraic correction expressions (``repair terms'')
    that allow the kernel to produce the correct result by the end of its execution.
  \end{enumerate*}
  In exchange for a small amount of computation and cache memory,
  the transformation reduces global memory transfers and improves data reuse.

  We present two instantiations of this paradigm: Rolling Update fusion and Split-K fusion.
  Both transformations are applicable and effective on attention-like operators.
  When applied to attention, rolling update fusion produces FlashAttention equivalents,
  suited for attention in prefill mode,
  and split-k fusion produces FlashDecoding equivalents for decoding attention.
  They are designed as primitives in schedule compilers and compose with other program
  transformations.

\item We present a compilation pipeline that integrates scheduling and tile-based optimizations.
  It separates low-level optimizations, which the tile optimizer handles,
  and high-level optimizations scheduling primitives,
  keeping scheduling concise and simplifying the search space.
  Scheduling applies to a loop-scalar intermediate representation (IR) that consists of
  loop nests and scalar expressions,
  which is well-suited for expression-manipulating transformations such as rolling update.
  After scheduling, \NAME{}'s translation engine lowers sections of loop-scalar IR to a
  tile IR, which can be optimized by the tile optimizer for high performance on device.
\end{itemize}

\noindent{}These two innovations represent key building blocks for supporting ---
\textit{natively within a compiler} ---
advanced algorithmic optimizations that produce high-performance tensor kernels.
Like Halide and TVM, \NAME{} takes as input a mathematical program definition and
optimization schedule,
but keeps the primitives focused on high-level optimizations.

We evaluate \NAME{} on 10 different attention-based operators from the literature and
several other operators, on multiple GPU architectures from NVIDIA and AMD.
Our results show that \NAME{}-generated kernels have lower latency than other
compiler-based frameworks, Triton~\cite{triton}, TVM~\cite{tvm},
FlexAttention~\cite{flexattn}, and Mirage~\cite{mirage},
including Triton-based implementations of FlashAttention.
Out of 320 optimized configurations (combinations of operators, sequence lengths, and GPUs),
\NAME{} generates the lowest-latency kernel in 284 cases.
\NAME{} achieves a speedup over the next best, already highly optimized alternative
of $1.35 \times$ (geomean of all cases),
up to $2.65 \times$ on Nvidia GPUs and up to $3.32 \times$ on AMD GPUs.
Further, in 101 out of 256 configurations, \NAME{} improves over the \mbox{state-of-the-art} kernels
from {CUTLASS~\cite{cutlass} based library implementations.}

\vspace{0.3em}
\noindent\textbf{Contributions:} The paper makes \mbox{several contributions:}
\begin{itemize}
\item We present \NAME{}, a tensor compiler that supports advanced loop fusion algorithms
  for reduction operators,
  while leveraging the advantages of both schedule-based and tile-based GPU optimization
  of \mbox{tensor programs.}
\item We present a novel paradigm for operator fusion that intentionally breaks data dependency
  and automatically derives algebraic repairs to obtain the correct result.
  Two instances, rolling update fusion and split-k fusion,
  are well-suited for optimizing attention-like operators.
\item We present a translation from scheduled loop programs to tile programs
  that a standard tile optimizer can process,
  freeing schedules from the burden of low-level optimizations.
\item We implemented Neptune on top of the Apache TVM schedule tensor compiler
  and the Triton tile tensor compiler.
  \NAME{}'s optimization approach is general and can be implemented over other similar
  tensor compilers. \NAME{} is available at {\url{https://github.com/uiuc-arc/neptune}}.
\item We thoroughly evaluated \NAME{} against state-of-the-art baselines. \NAME{}
  produces kernels that are faster
  than existing compilers by $1.35\times$ on average (geomean) and up to $3.32 \times$,
  on a diverse set of attention variants and four GPU architectures from Nvidia and AMD.
\end{itemize}
}

\section{Motivation}
\label{sec:motivation}
\newcommand{\nsum}[1]{\tagged{s_i}{#1}}
{\begin{figure*}[ht!]
  \centering\footnotesize
  \begin{subfigure}[b]{0.29\textwidth}
    \centering
    \begin{minted}[baselinestretch=0.93]{python}
# s_max:
for i in range(2):
  for j in range(4): # loop_j
    xmax[i] = max(
      xmax[i], inp[i, j])
# s_exp:
for i, j in grid(2, 4):
  xexp[i, j] = exp(
    inp[i, j] - xmax[i])
# s_sum:
for i, j in grid(2, 4):
  xsum[i] += xexp[i, j]
    \end{minted}
    \caption{The original program, consisting of 3 loop nests \smax, \sexp, and \ssum.}
    \label{fig:example-orig}
  \end{subfigure}
  \hspace{2pt}
  \begin{subfigure}[b]{0.32\textwidth}
    \centering
    \begin{minted}[baselinestretch=0.93]{python}
for i in range(2):
  for j in range(4):  # loop_j
    # s_max:
    xmax[i] = max(
      xmax[i], inp[i, j])
    # s_exp:
    xexp[i, j] = exp(       
      inp[i, j] - xmax[i])
    # s_sum:
    xsum[i] += xexp[i, j]
    \end{minted}
    \caption{
      Naive fusion of the original program.
      \sexp{} and \ssum{} are fused under \loopj{} without considering the data dependencies.
    }
    \label{fig:example-naive}
  \end{subfigure}
  \hspace{2pt}
  \begin{subfigure}[b]{0.36\textwidth}
    \centering
    \begin{minted}[baselinestretch=0.93]{python}
for i in range(2):
  xmax_0[i] = -inf
  for j in range(4):  # loop_j
    # s_max:
    xmax_1[i] = max(
      xmax_0[i], inp[i, j])
    # s_sum:
    xsum[i] = (
      exp(xmax_0[i] - xmax_1[i])
      * xsum[i]
      + exp(inp[i, j] - xmax_1[i]))
    xmax_0[i] = xmax_1[i]
    \end{minted}
    \caption{Correct fusion of Figure \ref{fig:example-orig}
      \textit{changes} the compute expressions of \ssum{}
      in addition to applying naive loop fusion.}
    \label{fig:example-ours}
  \end{subfigure}
  \hfill
  \begin{subfigure}[t]{0.4\textwidth}
    \centering
    \includegraphics[width=2.52\textwidth,trim=0mm 10mm 0mm 0mm,clip,page=1]{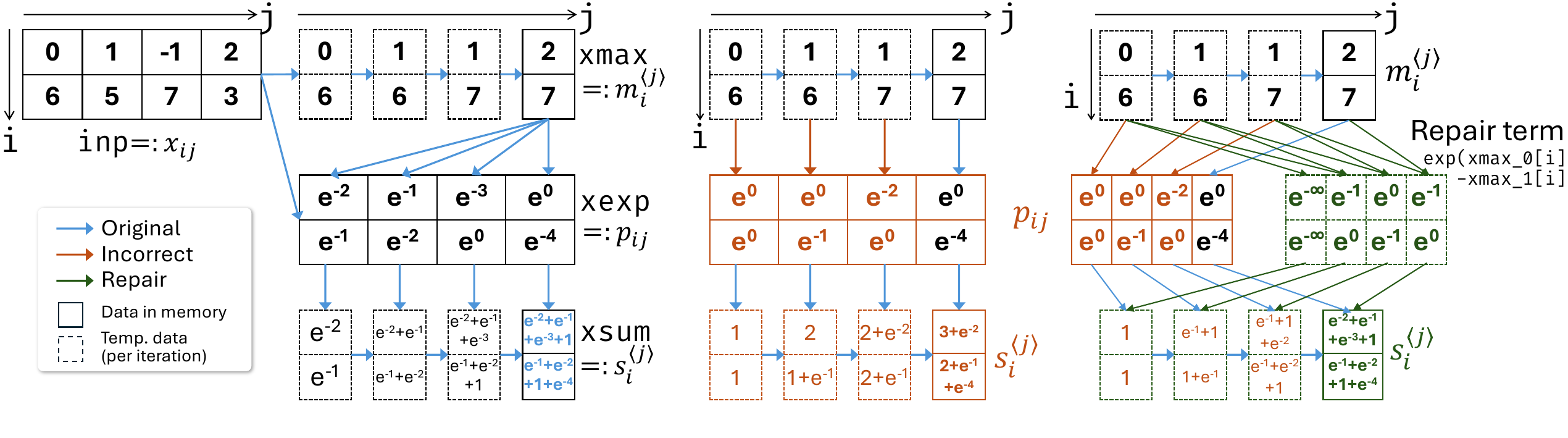}
    \caption{Materialized example of Figure \ref{fig:example-orig}.
      Data dependencies and values of this program are shown in \original{blue}.}
    \label{fig:idg-orig}
  \end{subfigure}
  \hspace{1pt}
  \begin{subfigure}[t]{0.24\textwidth}
    \centering
    \caption{Materialized example of Figure \ref{fig:example-naive}; \inp{} is omitted;
      incorrect results and dependencies in \incorrect{orange}.}
    \label{fig:idg-naive}
  \end{subfigure}
  \hspace{1pt}
  \begin{subfigure}[t]{0.33\textwidth}
    \centering
    \caption{Materialized example of Figure \ref{fig:example-ours}.
      Values of the repair term and the repaired \xsum{} are shown in \fixed{green}. Note,
      $\fixed{\nsum{3}} = \original{\nsum{3}}$.}
    \label{fig:idg-ours}
  \end{subfigure}
  \caption{An example program requiring reduction fusion, incorrect result using naive loop fusion,
    and a correctly fused program, shown on the first row.
    The second row (``materialized example'') shows the values of each matrix in the programs
    when applied to a concrete \inp{} matrix.
    \xmax{} and \xsum{} are each shown four times since their values vary over \texttt{j} iterations.}
  \label{fig:ru-example}
\end{figure*}
}
{\newcommand{\fsum}[1]{\fixed{\nsum{#1}}}
\newcommand{\fmax}[1]{\tagged{m_i}{\fixed{#1}}}
\newcommand{\ninp}[1]{x_{i#1}}
\newcommand{\nexp}[1]{p_{i#1}}
\newcommand{\nmax}[1]{\tagged{m_i}{#1}}

Figure \ref{fig:ru-example} shows an example program
where data dependencies prevent traditional fusion.
The input program (Figure~\ref{fig:example-orig})
runs on a 2-by-4 matrix \inp{} and consists of three loop nests:
a row-wise max \smax{}, an element-wise exponential \sexp{}, and a row-wise sum \ssum{}.

This program is similar to the softmax computation used in the attention
operator~\cite{attention}.
A major challenge in computing attention is that it does not scale to large sequence lengths ($L$),
because some intermediate tensors have size $L \times L$ and quadratic memory complexity.
Operator fusion can help avoid manifesting these tensors in memory,
but classic fusion techniques do not apply to this program due to its complex data dependency.
FlashAttention~\cite{flash-attention} is a specialized fusion solution for attention.
It enables fusion using a manually derived and proven algebraic correction term,
which is limited to attention.
\mbox{We now demonstrate how a correction term enables fusion for softmax.}

Figure \ref{fig:idg-orig} shows a concrete execution of the program \ref{fig:example-orig}
on an input \inp{}, where blue arrows mark data dependencies.
We shorthand matrix entries $\ninp{j} := \texttt{inp[i,j]}$ and $\nexp{j} :=
\texttt{\xexp[i,j]}$, and denote
$\nsum{j}$ and $\nmax{j}$ as the values of \xsumi{} and \xmaxi{} at iteration $j$.
\mbox{The program's final result~is~$\original{\nsum{3}}$.}

Fusing \ssum{} into \smax{}, both of which are reduction loops,
is an operator fusion that existing tensor compilers are not capable of.
If we disregard loop-carried dependencies and naively fuse \sexp{} and \ssum{} into
\smax{} with existing loop fusion, we obtain an incorrect program shown in Figure
\ref{fig:example-naive}.
The concrete execution in Figure \ref{fig:idg-naive} illustrates why it is incorrect:
fusing \sexp{} under \loopj{} changes some \sexp{} iterations to read from \xmax{} too early.
The brown arrows mark these incorrect data dependencies.
For example, in the first iteration, $\incorrect{\nexp{0}}$ reads from $\nmax{0}$
in the same iteration, instead of $\nmax{3}$ computed after 4 iterations.
Thus, $\incorrect{\nexp{0}}$ is incorrect and propagates to the final result $\incorrect{\nsum{3}}$.
We refer to this loop fusion without dependency checks as \textit{na\"ive fusion},
and current tensor compilers would reject this fusion in practice.

However, it is possible to build on na\"ive fusion and produce a correct program
if we \textit{update the compute expressions} of the fused loop nest \ssum{}.
We can find new expressions for \ssum{} that not only compute the current iteration,
but also \textit{repair} results from the previous iteration
as we move forward in the reduce dimension $j$.
Figure \ref{fig:example-ours} shows the correct fusion result based on this idea.
This program multiplies a \textit{repair term}
\texttt{exp(\xmaxprevi{} - \xmaxcurri)} with \xsumi{}.
Figure \ref{fig:idg-ours} shows the values of these repair terms
and the repaired $\fixed{\nsum{j}}$ values.
The repaired program is correct because $\fixed{\nsum{3}}$ is equal to $\original{\nsum{3}}$,
even though $\fixed{\nsum{j}} \neq \original{\nsum{j}}$ for $j < 3$.

To illustrate how the repair term repairs $\xsumi$ at every $j$ iteration,
we write out $\fixed{\nsum{j}}$ for three iterations,
as an expression of $\ninp{j}$ and $\nmax{j}$:

\begingroup
\small
\vspace*{-1em}
\begin{align*}
  \fsum{0} & = \exp(\ninp{0} - \fmax{0});
  \\
  \fsum{1} & = \exp(\nmax{0} - \nmax{1}) \cdot \fsum{0} + \exp(\ninp{1} - \nmax{1})
  \\
  & = \exp(\cancel{\nmax{0}} - \nmax{1}) \cdot \fixed{\exp(\ninp{0} - \cancel{\nmax{0}})}
  + \exp(\ninp{1} - \nmax{1}) \\
  & = \exp(\ninp{0} - \fmax{1}) + \exp(\ninp{1} - \fmax{1})
  \\
  \fsum{2} & = \exp(\cancel{\nmax{1}} - \nmax{2}) \cdot \left(\fixed{\exp(\ninp{0} -
      \cancel{\nmax{1}})
  + \exp(\ninp{1} - \cancel{\nmax{1}})}\right) + \exp(\ninp{2} - \nmax{2})
  \\
  & = \exp(\ninp{0} - \fmax{2}) + \exp(\ninp{1} - \fmax{2}) + \exp(\ninp{2} - \fmax{2})
  \numberthis\label{eq:example-result}
\end{align*}
\endgroup

We observe that the following behavior is key to a working repair term:
at each iteration $j$, multiplying the previous iteration result
$\tagged{s_i}{j - 1}$, which is an expression of $\tagged{m_i}{j - 1}$,
and replaces all uses of $\tagged{m_i}{j - 1}$ with $\tagged{m_i}{j}$.
If $\langle{j}\rangle$ is a tag on the array access, then this repair term is a
\textit{tag updater}.
We show in Section \ref{sec:tgo} that \NAME{} automatically finds
a tag updater for a large class of programs.

This solution to operator fusion adds a small amount of computational overhead,
as the repair term is evaluated once per iteration.
It also adds memory storage overhead to record the previous iteration results
(i.e. \xmaxprev{} and \xmaxcurr{}) in GPU registers or shared memory.
This memory overhead is low, as reduction output arrays are small
compared to other arrays in the program, such as \xexp{} and \inp{}.
Fusion benefits, including reduced global memory transfers and improved data reuse,
often far outweigh the overhead, as evidenced by FlashAttention's performance improvements.
}

\section{System Overview}
\label{sec:design}
{\begin{figure}[ht]
  \centering
  \includegraphics[width=0.7\columnwidth,page=1]{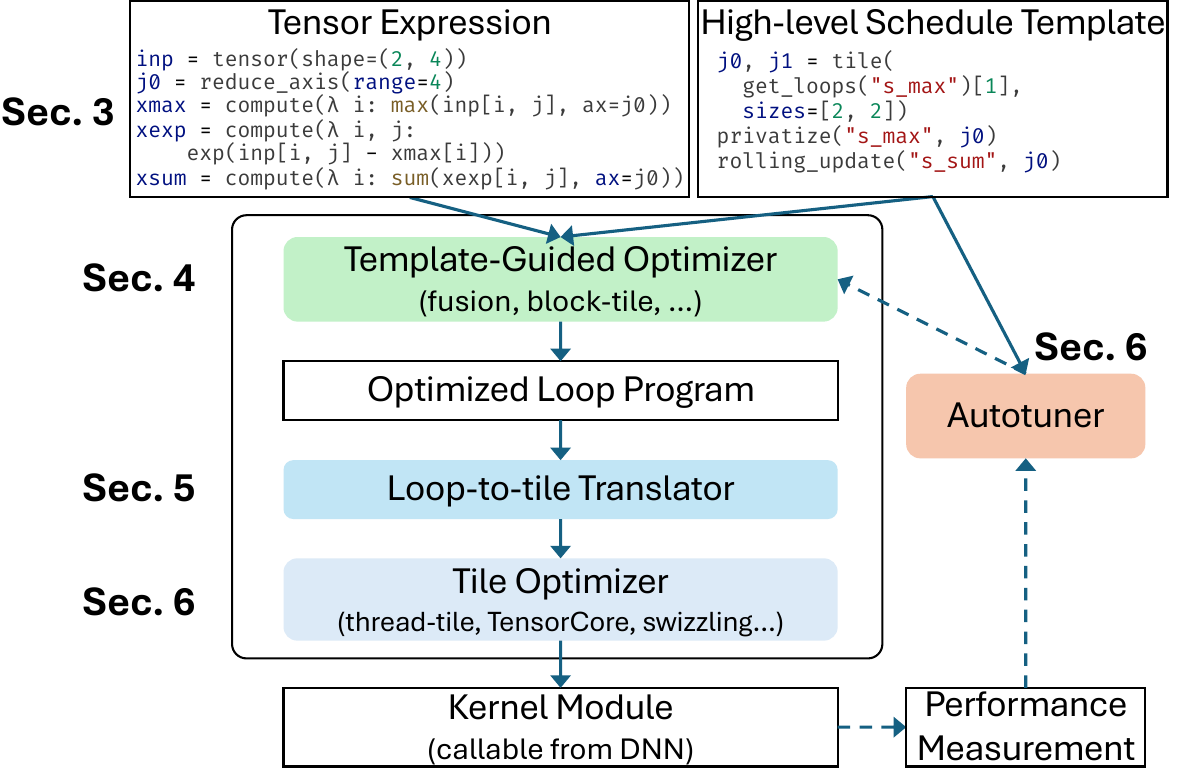}
  \caption{The main components of \NAME{}.}
  \label{fig:system-overview}
\end{figure}

Figure \ref{fig:system-overview} summarizes the main components of \NAME{}.
\NAME{} takes as input a program written in tensor expressions and a schedule template
for the program.
The schedule template is a user-provided recipe with a list of \textit{high-level
optimizations} for the program.
\NAME{} optimizes the input program with the following steps:

\begin{itemize}[leftmargin=*]
  \item The template-guided optimizer converts the tensor expression into
    a program in \NAME{}'s loop-scalar IR,
    and applies the transformations in the template on it.
    \NAME{}'s two novel advanced fusion algorithms (\S\ref{sec:tgo}) are in this optimizer,
    and they compose with other standard transformations.
    This step outputs a loop program that has undergone high-level optimizations.
  \item The loop-to-tile translator converts the loop program into a program in \NAME{}'s tile IR.
    This translation algorithm (\S\ref{sec:l2t}) is automatic and covers a majority of programs
    representable in the loop-scalar IR.
    The tile optimizer applies tile-specific transformations to the program
    (e.g., data movement, tile data layout, and HW-specific execution)
    and produces an executable kernel.
  \item \NAME{}'s autotuner (\S\ref{sec:impl}) mutates the schedule template and
    optimization parameters
    to find a better list of transformations that further improves kernel performance.
\end{itemize}

\textbf{Input: Tensor Expression.}
\NAME{}'s input programs are written in a compact tensor expression language which
translates straightforwardly into \NAME{}'s loop-scalar IR.
Figure \ref{fig:system-overview} (top left) shows the tensor expressions
that produce the loop program in Figure \ref{fig:example-orig}.
Tensor expressions are written in TVM expression language here;
other tensor expression languages such as JAX's jaxpr language could be used in principle.

\textbf{Input: Schedule Template.}
A schedule template is a list of high-level optimization steps
the user writes to describe how to optimize the program.
Each step calls a transformation primitive in \NAME{} with some parameters.
The template-guided optimizer applies each step in the template to the program.
We describe the optimizer and the transformation primitives available in Section \ref{sec:design}.
Figure \ref{fig:system-overview} (top right) shows an example template
that is applicable to the given program.

While schedule templates resemble the scheduling languages
used in some tensor compilers, \NAME{} templates are concise.
In a few lines of code, a user can express all the transformations
needed to optimize a complex operator with multiple loop nests.
Schedule templates only describe high-level optimizations
that benefit from user direction,
while \NAME{} delegates low-level optimizations to its tile optimizer without user intervention.
An example of tensor expression and schedule for \mbox{attention is in
Appendix~\appref{sec:app:neptune-inputs}\appcite{}.}

\def\ZZ{\mathbb{Z}}
\newcommand{\loopnest}[3]{nest(#1:#2)\{ #3 \}}
\NewDocumentCommand{\optsub}{m g}{#1\IfValueT{#2}{_{#2}}}
\NewDocumentCommand{\xacc}{g}{\optsub{X}{#1}[\optsub{\accf}{#1}(\bi)]}

\subsection{Preliminaries and Notation}

\textbf{Loop Nest Notation.}
We use a compact notation for loop nests: $\loopnest{\bi}{\bn}{s}$
to represent $K$ perfectly nested loops with upper bounds $\bn := (n_1, \dots, n_K)$ and body $s$.
\NAME{} normalizes the lower bound of loops to $0$ without loss of generality.
$\bi \in \ZZ^K$ is an iteration vector of the nest.
The iteration domain $\dom{s}$ is the set of values $\bi$ can take, determined by $\bn$.
We denote the previous iteration of $\bi$ by $\iprev{\bi}$
and the next iteration by $\inext{\bi}$.

\textbf{Affine Access Function and Tags.}
In a tensor access $\xacc$ under a loop nest (with iteration vector $\bi$),
$\accf$ is an \textit{affine access function} that projects $\bi$ to tensor indices.
The tensor access \xsumi{} under loops $i$ and $j$ has an access function $\accf(i, j) = i$
(example from Figure \ref{fig:example-orig}).
When a tensor entry is updated multiple times,
we add a \textit{tag} on the access to mark the iteration we refer to:
$\tagged{\xacc}{j}$ ($\tagged{\xsumi}{j}$ for example).
Tags reflect temporal data dependency on the same tensor entry.
We have used tags in Figure \ref{fig:ru-example}.

\textbf{Reduction in \NAME{} IR.}
A reduce loop nest is a loop nest that expresses a reduction operation.
It consists of map loops and reduce loops,
where the output is only indexed by the map loop variables,
and each output value is updated as many times as the reduce loops execute.
A reduce loop nest in \NAME{}'s loop-scalar IR has the form
\begin{equation}
  \rnest{\bi}{\bn}{\bj}{\bm}{\ \xacc = \xacc \cf expr(\bi, \bj)\ } \label{eq:reduce-iter-form}
\end{equation}
The notation $\rnest{\bi}{\bn}{\bj}{\bm}{s}$ distinguishes reduce loops from map loops.
$f$ is the reducer, an associative binary function, and $a \cf b$ is equivalent to $f(a, b)$.

\textbf{Recurrent and Explicit Forms of Reductions.}
Equation \ref{eq:reduce-iter-form} is the \textit{recurrent form} of a reduction,
which iteratively updates the tensor $X$.
We define the reduce operator $\fold$ to express the value of $X$ compactly:
$\fold(f, 0 \leq j \leq j_0, g(j))$ applies $f$ to fold over $g(j)$
from $j = 0$ to $j_0$ to produce a single value.
For instance, $\fold(+, 0 \leq j \leq 2, \xexp[i, j])$ equals
$\xexp[i, 0] + \xexp[i, 1] + \xexp[i, 2]$.
We refer to this $\fold$-based expression as the \textit{explicit form}
for the value of $X$.
}

\section{Template-Guided Optimizer}
\label{sec:tgo}
\def\redpred{\mathcal{L}_r}
{\def\RR{\mathbb{R}}
\def\calx{\mathcal{X}}
\newcommand{\bs}[1]{\boldsymbol{#1}}

\newcommand{\fsum}[1]{\fixed{\nsum{#1}}}
\newcommand{\fmax}[1]{\tagged{m_i}{\fixed{#1}}}
\def\xri{\access{X_r}{r}{\bi}}
\def\xti{\access{X_t}{t}{\bi}}
\def\cij{\access{C}{c}{\bi, \bj}}
\def\cijp{\access{C}{c}{\bi, \bj'}}
\newcommand{\foldbj}[2]{\fold\left(f, 0 \preceq \bj' \preceq #1, #2\right)}

\def\imrr{\textsc{InlineMapReturnReduce}}
\def\match{\textsc{MatchReducePattern}}

The template-guided optimizer translates the input tensor expression program into
loop-scalar IR.
This translation is a standard process, as the tensor expression language
and the loop-scalar IR are both similar to those in many tensor compilers.
The abstract syntax of this IR is shown in Figure \ref{fig:loop-ir-syntax} of the next section.
Next, the template-guided optimizer applies the transformations in the template to the
loop IR program.
The template is a list of instructions that refer to transformation primitives in \NAME{},
including standard loop transforms, data layout transforms,
and data placement transforms (e.g., caching in shared memory).
\NAME{} also provides novel transformations that enable reduction fusion, described next.

\subsection{Algebraic Correction Analysis for Reduction Fusion}
\label{sec:repair-function}

\NAME{} is capable of fusing reduction operations.
In Figure \ref{fig:ru-example}, we have manually fused the reductions \ssum{} and \smax{}
and fixed the result with a repair term.
Finding this repair term is the main challenge of reduction fusion.
To enable generalized and reusable reduction fusion in a compiler,
\NAME{} has a novel analysis that derives a repair term automatically from the program,
which is the core behind its reduction fusion primitives.
We give an intuition of how \NAME{} finds the repair term
from the motivating example (\S\ref{sec:motivation}) and show the formal generalization here.

\textbf{Characterizing a Reduction.}
\NAME{} implements generalized fusion of a reduction loop (a consumer reduction) that
depends on values produced by one or more other reduction loops (producer reductions). We
consider consumer loops that decompose into a reducer function $f$ performing reduction
over terms and a function $g$ that produces those terms:
\[
  f: \calx \times \calx \to \calx\ (\text{associative}); \qquad
  g: \calx \times \calx \to \calx.\ g(r, c)
\]
In $g(r, c)$, $r$ is the result of the producers,
and $c$ is a collection of values not produced by a reduction (e.g., a program input).
For simplicity, this section will provide intuition for a single producer reduction,
which will be formalized for multiple producers in section \ref{sec:rolling-update}.

\textbf{Reduction Repair Function.}
We take inspiration from the repair term \(\exp(\tagged{m_i}{j - 1} - \tagged{m_i}{j})\)
in the example, and generalize it to a \textit{repair function}:
\[
  h: \calx \times \calx \times \calx \to \calx.\quad h(t, r, r')
\]
This function takes three arguments: $t$, an in-progress reduction value from the
consumer reduction (\ssum), and $r$ and $r'$, two reduction values from different
iterations of the producer reduction (\smax). Like the repair term, we aim to find a
specific $h$ capable of taking a reduction value $t$ computed with $r$,
and repair it to a reduction value $t'$ computed with $r'$. For a reduction characterized
by $f$ and $g$ above,
satisfying two conditions allows $h$ to work as a repair function:
\begin{align}
  h\left(g(r, c), r, r'\right) & = g(r', c)                    & \forall r, r', c \in
  \calx    & \label{eq:h-conditions-1} \\
  h\left(x \cf y, r, r'\right) & = h(x, r, r') \cf h(y, r, r') & \forall x, y, r, r' \in
  \calx & \label{eq:h-conditions-2}
\end{align}
Namely, $h$ replaces the $r$ argument of $g$, and $h$ distributes with the reducer $f$.
Importantly, an $h$ satisfying condition \ref{eq:h-conditions-1} can be mechanically constructed
if $g(r, c)$ is invertible in the argument $c$ (called the inverse function $g_c^{-1}$):
\begin{equation}
  h(t, r, r') = g(r', g_c^{-1}(r, t)) \label{eq:h-solution}
\end{equation}

Section \ref{sec:rolling-update} presents the full formal arguments that Equations
\ref{eq:h-conditions-1}-\ref{eq:h-solution} are correct,
but intuitively they allow out-of-date $r$-values $r_0$ in a reduction to be replaced
with new values $r_1$.
For example, consider a small two-term reduction $g(r_0, x) \cf g(r_0, y)$ which
transforms through $h$ to $ g(r_1, x) \cf g(r_1, y)$:
\begin{align*}
  h(g(r_0, x) \cf g(r_0, y), r_0, r_1) & = h(g(r_0, x), r_0, r_1) \cf h(g(r_0, y), r_0,
  r_1) & \text{Using Eq. \ref{eq:h-conditions-2}} \\
  & = g(r_1, x) \cf g(r_1, y)                           & \text{Using Eq. \ref{eq:h-conditions-1}}
\end{align*}

Consider again Equation \ref{eq:example-result}. Our general formulation maps to this example with:
\begin{align*}
  f(t_0, t_1)                              & = t_0 + t_1
  &  & \text{Reducer Function} \\
  g(m_i, x_{ij})                           & = \exp(x_{ij} - m_i)
  &  & \text{Reduction Terms}  \\
  h(t, \tagged{m_i}{k}, \tagged{m_i}{k+1}) & = t \exp( \tagged{m_i}{k} -
  \tagged{m_i}{k+1}) &  & \text{Repair Term}
\end{align*}
The functions $f$ and $g$ are derived from the program, while $h$ simplifies from
Eq.~\ref{eq:h-solution}.
Algebraic correction analysis identified appropriate $f$, $g$, and $h$,
and enables multiple operator fusion transformations in \NAME{},
described in sections \ref{sec:rolling-update} and \ref{sec:split-k}.

\subsection{Rolling Update}
\label{sec:rolling-update}

Rolling update is \NAME{}'s novel transformation that fuses reduction loop nests.
It applies to a reduce loop nest $L_t$ (the consuming reduction/fusee)
and a target loop $l$ (the \textit{rolling loop}) to fuse $L_t$ under:
\texttt{RollingUpdate($L_t$,$l$)}.
Producing reductions are not given as arguments and are discovered via data dependency analysis.
Rolling update requires that the fusee $L_t$ is a reduce loop nest,
and that the rolling loop $l$ does not enclose $L_t$.

\begin{wrapfigure}{r}{0.5\columnwidth}
  \vspace{-\baselineskip}
  \begin{algorithm}[H]
    \small
    \caption{\textsc{RollingUpdate}($L_t$, $l$)}
    \label{alg:rolling-update}
    \KwIn{$L_t$: the target loop nest to be fused. $L_t$ holds a reference to the entire program.}
    \KwIn{$l$: the loop to fuse $L_t$ under (``rolling loop'').}
    \KwOut{The fused loop nest}
    $\redpred$ = \imrr{}($L_t$)\; \label{alg:ru:df}
    \For{$L \in \redpred \cup \text{set}(L_t)$}{  \label{alg:ru:loop-tr-start}
      $L$ = \textsc{NaiveLoopFusion}($L$, $l$)\;
    }\label{alg:ru:loop-tr-end}
    $(f, g)$ = \match($L_t$, $\redpred$)\; \label{alg:ru:patmatch}
    $h$ = \textsc{SolveRepairFuncH}($g$)\; \label{alg:ru:solve}
    \textsc{ValidateHCommutative}($h$, $f$)\; \label{alg:ru:validate}
    $M_x$ = dict()\; \label{alg:ru:apply-start}
    \For{$L \in \redpred$}{ \label{alg:ru:cache-start}
      $(x_{\text{prev}}, x_{\text{curr}})$ = \textsc{CacheReducePrevResult}($L$)\;
      $M_x[L]$ = $(x_{\text{prev}}, x_{\text{curr}})$\;
    } \label{alg:ru:cache-end}
    $L_t$ = \textsc{ApplyRepairTerm}($L_t$, $h$, $M_x$)\; \label{alg:ru:apply-end}
    \Return{$L_t$}
  \end{algorithm}
\end{wrapfigure}

Algorithm \ref{alg:rolling-update} outlines the full rolling-update algorithm.
There are four major steps in rolling update, which we explain in the following paragraphs.
When any step fails, such as precondition check, pattern matching, or repair function solving,
rolling update returns the original program.
A step-by-step example of applying rolling update to Figure \ref{fig:example-orig}
is in Appendix \appref{sec:app:step-by-step}.

\textbf{Step 1: Dataflow Reorganization.}
Rolling update performs an analysis on the dataflow graph
to find the producing reductions of $L_t$ as a set $\redpred$ (algorithm line \ref{alg:ru:df}).
Then on each path from $L_t$ that ends in $\redpred$ (excluding both ends),
rolling update inlines all the loop nests, similar to how we inlined \sexp{} in the example.
These inlined loop nests are predecessors of $L_t$,
so their computation eventually accumulates in $L_t$.
After inlining, all of $\redpred$ are immediate producers of $L_t$.

\textbf{Step 2: Loop Transformation.}
Rolling update fuses $\redpred$, followed by $L_t$, under $l$ using naive loop fusion
(line \ref{alg:ru:loop-tr-end}), skipping loop nests that are already under $l$.
This step ensures that all loop nests in $\redpred$ share an outer loop nest with $L_t$.
$L_t$ is now in fused position and produces incorrect results, which need to be repaired.

\textbf{Step 3: Finding the Repair Function.}
Rolling update uses Equations \ref{eq:h-conditions-2} and \ref{eq:h-solution}
(\S\ref{sec:repair-function}) to find the repair function $h$.
It first needs to understand the structure of the $L_t$ loop nest.
On algorithm line \ref{alg:ru:patmatch}, it matches $L_t$
to the structure of a reduction loop nest in \NAME{}'s IR.
This step extracts $f$ and $g$ as expressions from the input program:
\begin{equation}
  \rnest{\bi}{\bn}{\bj}{\bm}{\xti = \xti \cf{} g\left(\xri, \cij\right)} \label{eq:ru-pattern}
\end{equation}
This pattern matches a reduction in \NAME{}'s IR (Eq.~\ref{eq:reduce-iter-form}),
with additional structure $g(\dots)$ on the right of the reducer $f$.
$X_t$ is the output of $L_t$,
$X_r$ is the output of the producing reduction of $L_t$ (which is $L_r \in \redpred$),
and $C$ is non-reduce input to $L_t$ (not produced by a reduction, e.g., program input).
We use one producing loop as an example for simplicity and will generalize it later.

Given $f$ and $g$, rolling update solves Eq.~\ref{eq:h-solution}
to find the repair function $h$ (line \ref{alg:ru:solve}),
and proves if $h$ satisfies Eq.~\ref{eq:h-conditions-2} (line \ref{alg:ru:validate}),
both using a symbolic solver.

\textbf{Step 4: Applying the Repair Function.}
Rolling update applies the repair function $h$ to the body of $L_t$,
performing the rewrite from Eq.~\ref{eq:ru-pattern} to the following (loop nest omitted):
\begin{equation}
  \xti = h\left(\xti, \tagged{\xri}{\prevj}, \tagged{\xri}{\bj}\right)
  \cf{}\ g\left(\xri, \cij\right) \numberthis \label{eq:fixed-recurrent}
\end{equation}
This rewrite corresponds to the example in Figure \ref{fig:example-ours},
where we applied the repair term on the left-hand side of the reducer ($+$)
while not changing the right-hand side.

Eq.~\ref{eq:fixed-recurrent} requires both $\tagged{\xri}{\prevj}$ and $\tagged{\xri}{\bj}$
(output of $L_r$ at the previous and current iterations),
but a reduction only keeps the output of the current iteration.
Therefore, we need to transform the producing reduction to retain the results of two iterations.
In the example (Figure \ref{fig:example-ours}),
\smax{} is transformed to produce \xmaxprev{} and \xmaxcurr{}, one for each iteration.
Lines \ref{alg:ru:cache-start}-\ref{alg:ru:cache-end} describe this process:
rolling update transforms each $L_i \in \redpred$ to produce two output tensors.
Finally, line \ref{alg:ru:apply-end} applies the above rewrite
and concludes the rolling update algorithm.

\textbf{Generalizing to Multiple Producing Reductions.}
Rolling update generalizes to a variety of dataflow patterns,
including multiple producing reductions and non-reduce inputs for the fusee $L_t$,
which we omitted in the repair function analysis for simplicity.
Now, we discuss how to handle multiple producing reductions
$L_{r1}, \dots, L_{rN}$ and non-reduce inputs $C_1, \dots, C_M$.

We make the following changes to the previous definitions in algebraic correction
analysis (\S\ref{sec:repair-function}):
$g$ is a function of $\calx^N \times \calx^M \to \calx$,
which we denote as $g(\bs{r}, \bs{c})$.
The repair function $h: \calx \times \calx^{N} \times \calx^{N} \to \calx$
now reads the previous and current iteration results from each of the $N$ producing reductions:
$h(t, \bs{r}, \bs{r}')$.
We retain the conditions and solution for $h$ from Section \ref{sec:repair-function},
which remains formally similar, with $r$, $r'$ and $c$ becoming vectors:
\begin{align}
  h\left(g(\bs{r}, \bs{c}), \bs{r}, \bs{r}'\right)    & = g(\bs{r'}, \bs{c})
  \ \ \Rightarrow \ \  h(t, \bs{r}, \bs{r}')    = g(\bs{r'}, g_c^{-1}(\bs{r}, t)) &
  \forall \bs{r}, \bs{r}' \in \calx^N, \bs{c} \in \calx^M \label{eq:h-cond-1-general}
  \\
  h\left(x \cf y, \bs{r}, \bs{r}'\right)              & = h(x, \bs{r}, \bs{r}') \cf h(y,
  \bs{r}, \bs{r}')                                                     &
  \forall x, y \in \calx, \bs{r}, \bs{r}' \in \calx^N \label{eq:h-cond-2-general} &
\end{align}

\textbf{Support for Non-Invertible Functions.}
\NAME{} can find a repair term for a class of compute patterns even when
the $g$ function is non-invertible.
If $g$ is not invertible against $\bs{c}$,
making a change of variables $C := g_c(\bs{c})$ may help the solver find a solution:
\[
  \exists g_c: \calx^M \to \calx, g_0: \calx^N \times \calx \to \calx
  \quad \text{s.t.} \quad
  g(\bs{r}, \bs{c}) \equiv g_0(\bs{r}, g_c(\bs{c}))
\]
If this change of variables produces a function $g_0(\bs{r}, C)$ that is invertible in $C$,
\NAME{} can still find the repair function $h$.
We denote the inverse function of $g_0(\bs{r}, C)$ as $g_0^{-1}(\bs{r}, t) = C$.
In this case, the solution to the repair function $h$ in Equation \ref{eq:h-solution}
still holds (replacing $g$ with $g_0$):
\[
  h(t, \bs{r}, \bs{r'}) = g_0(\bs{r'}, g_0^{-1}(\bs{r}, t))
\]

An example where this change of variables is useful is
$g(r, c) = \exp(r - \mathrm{relu}^2(c))$.
This function is not invertible in $c$ because it applies the non-injective Squared ReLU function.
\NAME{}'s solver finds the variable substitution $C = \mathrm{relu}^2(c)$,
and solves the alternative problem $g(r, C) = \exp(r - C)$ which is invertible.

\textbf{Correctness of the Repaired Program.}
\renewcommand{\foldbj}[2]{\fold\left(f, 0 \preceq \bj \preceq #1, #2\right)}
A correct program is one where the $L_t$ loop nest after rolling update
produces the same result as in the original program.
\begin{theorem}[Correctness of the Repaired Program]
  \label{thm:rollup-correct}
  If rolling update succeeds and rewrites the reduce loop nest $L_t$
  to Eq.~\ref{eq:fixed-recurrent},
  the value of $X_t$ in the repaired program is the same as in the original program
  after execution of $L_t$.
\end{theorem}
We sketch a proof for this theorem here;
the full proof of this theorem and the lemmas used are in Appendix \appref{sec:app:proofs}.
The loop nest before fusion is shown as a program in Eq.~\ref{eq:ru-pattern},
which iteratively updates $X_t$ for iterations $\bj$ from $0$ to $\bm$.
We convert this program into an expression for the value of $X_t$,
expressed using the \textit{explicit form} of a reduction:
\begin{equation}
  \original{\tagged{\xti}{\bm}} = \foldbj{\bm}{g\left(\xri, \cij\right)}
  \label{eq:original-explicit}
\end{equation}

To prove the correctness of rolling update,
we show that the loop nest after fusion, which is given by Eq.~\ref{eq:fixed-recurrent},
also corresponds to the same $X_t$ value as Eq.~\ref{eq:original-explicit}.
This calculation in the full proof uses the repairing property of $h$ to simplify the expression,
so we formally define it, which we refer to as the \textit{tag-updating property}:
\def\jf{\incorrect{\bj_f}}
\def\jt{\fixed{\bj_t}}
\begin{definition}
  \label{def:h-tag-update}
  A function $h$ \textit{tag-updates} a reduction characterized by reducer $f$ and function $g$
  if it satisfies the following condition:
  \begin{align*}
    & h \left(
      \foldbj{\bj_0}{g\left(\tagged{\xri}{\jf}, \cijp\right)},
      \tagged{\xri}{\jf},
      \tagged{\xri}{\jt}
    \right)                                                       \\
    =\  & \foldbj{\bj_0}{g\left(\tagged{\xri}{\jt}, \cijp\right)}
    \qquad \forall \bj_0, \jf, \jt \in \dom{s}, \jf \preceq \jt \preceq \bj_0
  \end{align*}
  \vspace{-1.5em}
\end{definition}
This definition matches the intuition that $h$ replaces out-of-date $r$-values
(ones with tag $\jf$) in a reduction with new values (ones with tag $\jt$).
Section \ref{sec:repair-function} provided a solution for the repair function $h$
(Equations \ref{eq:h-conditions-1}-\ref{eq:h-solution}).
We check that the solution satisfies this definition of the tag-updating property with a lemma:
\begin{lemma}
  \label{lemma:h-tag-update}
  If $h$ satisfies Equations \ref{eq:h-conditions-1}-\ref{eq:h-solution},
  then it tag-updates as Def.~\ref{def:h-tag-update} requires.
\end{lemma}

We prove Lemma \ref{lemma:h-tag-update} and Theorem \ref{thm:rollup-correct}
in Appendix~\appref{sec:app:proofs}.
Their proofs use the associative property of the reducer $f$,
which is true for real ($\RR$) inputs.
In practice, rolling update applies to programs over floating-point inputs,
for which $f$ is not strictly associative.
We empirically check the numerical accuracy of the transformed program
(evaluation in Appendix \appref{sec:app:num-stab})
and find it to be highly accurate.

\textbf{Tradeoffs and Scope of Rolling Update.}
Rolling update implements reduction fusion. It extends operator fusion
to a mix of elementwise and reduction operations, bringing the benefits of operator fusion:
improving data reuse, reducing global memory transfers, and reducing peak memory usage.
On the other hand, rolling update adds extra computation to the target loop nest $L_t$
due to the repair term $h$.
Moreover, since the new program maintains the result of last and current iterations,
the rolling loop $l$ becomes harder to parallelize due to more complex data flow.
We describe another fusion approach in Section~\ref{sec:split-k} that alleviates the second issue.

There exist cases where \NAME{}'s solver truly fails to find a solution even with the
change of variables technique.
Such failures often indicate that fusion is inherently difficult between the operators.
One example is the simultaneous computation of mean and variance
known as the Welford algorithm \cite{welford}.
For numerical stability, it is standard to compute the variance of data
by first shifting the data by their mean:
$s(\bs{x}) = \sum_i (x_i - m)^2 / (N-1)$
where $m(\bs{x}) = \sum_i x_i / N$ and $\bs{x} \in \mathbb{R}^N$.
The Welford algorithm provides a formula $f$ to update the variance online
for every incoming sample: $s_{i + 1} = f(s_i, x_{i + 1})$.
\NAME{}'s rolling update is applicable to fusing the mean and variance computation
and in principle can produce a fused one-pass algorithm similar to Welford's algorithm,
but the algebraic correction analysis fails to find a solution.
\NAME{}'s solver currently does not take into consideration that
the previous/current mean, the current variance, and the incoming sample
are all related quantities,
which can further simplify the compute expressions.

\subsection{Split-K Update}
\label{sec:split-k}

\def\sml{\texttt{s\_max\_local}}
\def\ssl{\texttt{s\_sum\_local}}
\def\smg{\texttt{s\_max\_global}}
\def\ssg{\texttt{s\_sum\_global}}

\NAME{} is capable of another reduction fusion transformation, \textit{split-k update}.
While rolling update makes the rolling loop harder to parallelize,
split-k update avoids this issue and aims to maximize parallelism.
Split-k breaks dependencies of reductions so that \emph{partial reductions} are computed in parallel
in one loop nest, and a second loop nest combines and repairs the partial results simultaneously.
Applying split-k update to attention produces kernels similar to
FlashDecoding~\cite{flash-decoding}.
This transformation is named because it resembles the split-k strategy in matrix
multiplication~\cite{splitk}.

Split-k update derives from the same break-and-repair paradigm as rolling update,
using the same analysis and repair function $h$.
It also reuses dataflow reorganization (step one) of rolling update.
The difference lies in the values applied to $h$:
while rolling update uses $h$ to tag-update by one iteration,
split-k update uses $h$ to replace local reduction results with global reduction results.

Figure \ref{fig:su-result} shows the result of applying split-k update
to the motivating example in Figure \ref{fig:example-orig},
An accompanying data dependency graph is shown in Figure \ref{fig:su-result-ddg}.
The fusee \ssum{} and the producing reduction \smax{} are each split into a local reduce loop nest
(\sml{}, \ssl{}) and a global reduce loop nest (\smg{}, \ssg{}).
The local nests compute partial reductions, and the global nests combine
partial results.
The repair function $h$ is applied to the global nest of the fusee, \ssg{}.
Unlike rolling update, the ``rolling'' loop $j0$ is now highly parallelizable.

\newcommand{\loc}[1]{#1^\text{local}}
\newcommand{\glb}[1]{#1^\text{global}}  

\begin{wrapfigure}{r}{0.52\columnwidth}
  \begin{algorithm}[H]
    \small
    \caption{\textsc{SplitKUpdate}($L_t$, $l$)}
    \label{alg:splitk-update}
    \KwIn{$L_t$: the target loop nest to be fused}
    \KwIn{$l$: the target loop to fuse $L_t$ under}
    \KwOut{A pair of loop nests: the local and global versions of fused $L_t$}
    $\redpred$ = \imrr{}($L_t$)\;
    $(\loc{M}, \glb{M})$ = (dict(), dict())\; \label{alg:su:pr-start}
    \For{$L_r \in \redpred$}{
      $(\loc{L_r}, \glb{L_r})$ = \textsc{FuseAndPrivatize}($L_r$, $l$)\;
      $(\loc{M}[L_r], \glb{M}[L_r])$ = $(\loc{L_r}, \glb{L_r})$\;
    } \label{alg:su:pr-end}
    $L_t$ = \textsc{ReplaceTensorReads}($L_t$, $\loc{M}$)\; \label{alg:su:replace}
    $(\loc{L_t}, \glb{L_t})$ = \textsc{FuseAndPrivatize}($L_t$, $l$)\; \label{alg:su:pr-l0}
    $h$ = \textsc{RollingUpdateSolveRepairFunc}($\loc{L_t}, \redpred$)\; \label{alg:su:h}
    $\glb{L_t}$ = \textsc{SplitKApplyRepairTerm}($\glb{L_t}$, $h$, $\loc{M}$)\; \label{alg:su:apply}
    \Return{$(\loc{L_t}, \glb{L_t})$}\;
  \end{algorithm}
\end{wrapfigure}

\textbf{Reduction Privatization.}
Splitting a reduction into a partial and a global reduction
is a standard optimization technique called reduction privatization,
which \NAME{} provides as a transformation primitive.
It applies to a reduce nest $L$ that runs $K$ reduce iterations and creates two reduce nests:
an $\loc{L}$ with $K$ reduce iterations distributed to $k$ threads,
and an $\glb{L}$ with $\lceil K / k \rceil$ reduce iterations to combine the partial results.
Privatization used in Figure \ref{fig:su-result} \mbox{has $K = 4$ and $k = 2$.}

Privatization requires a tile size $k$ as input argument. Split-k uses a combined transformation
of naive loop fusion followed by privatization: \texttt{FuseAndPrivatize},
as naive loop fusion produces a loop nest structure that privatization can infer the tile
size $k$ from.

\begin{figure}[t]
  \centering
  \begin{adjustbox}{valign=b,minipage={.6\textwidth}}
    \small
    \begin{minted}{python}
for i, j0 in grid(2, 2):  # <- "rolling" loop j0
  for j1 in range(2): # s_max_local
    max_l[i, j0] = max(max_l[i, j0], inputs[i, j0 * 2 + j1])
  for j1 in range(2): # s_sum_local
    sum_l[i, j0] += exp(
      inputs[i, j0 * 2 + j1] - max_l[i, j0])
for i, j0 in grid(2, 2): # s_max_global
  max_g[i] = max(max_g[i], max_l[i, j0])
for i, j0 in grid(2, 2): # s_sum_global
  sum_g[i] += exp(
    max_l[i, j0] - max_g[i]) * sum_l[i, j0]
  \end{minted}
    \caption{The result of applying split-k update to Fig.~\ref{fig:example-orig}.
      Split-k update privatizes two reductions in the original program,
    creating a local and a global reduction for each.}
    \label{fig:su-result}
  \end{adjustbox}\hspace{3pt}
  \begin{adjustbox}{valign=b,minipage={.35\textwidth}}
    \centering\includegraphics[width=0.9\textwidth,trim=0mm 15mm 90mm
    2mm,clip,page=2]{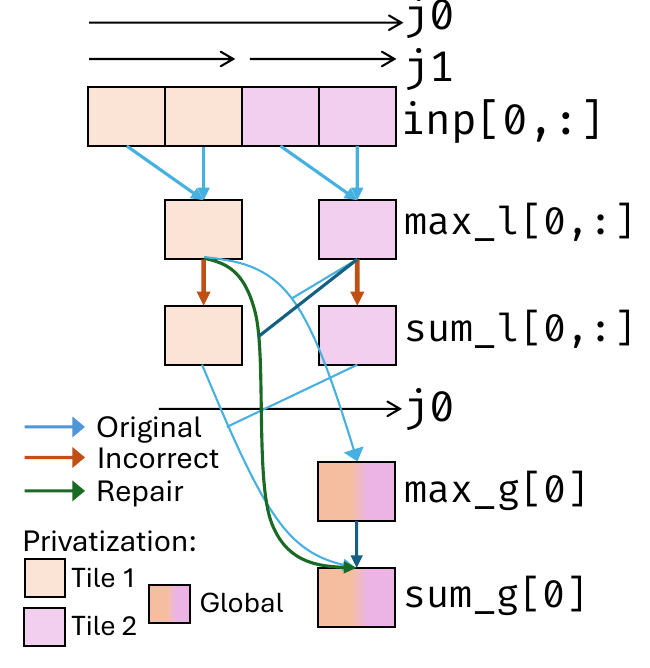}
    \caption{Data dependency graph of split-k transformed program in Figure \ref{fig:su-result}.}
    \label{fig:su-result-ddg}
  \end{adjustbox}
\end{figure}

\textbf{Split-K Algorithm.}
Algorithm \ref{alg:splitk-update} outlines the split-k update algorithm.
We explain this algorithm in comparison to rolling update:
\begin{itemize}
  \item Split-k update applies dataflow reorganization of rolling update to get $\redpred$.
  \item On algorithm lines \ref{alg:su:pr-start} to \ref{alg:su:pr-end},
    split-k update uses \texttt{FuseAndPrivatize}
    to fuse all producing reductions of $L_t$ under $l$,
    and tracks their local and global loop nests $\loc{L}$ and $\glb{L}$.
  \item Line \ref{alg:su:replace} replaces tensor reads in $L_t$
    such that it reads from the local nests of the producing reductions.
    Line \ref{alg:su:pr-l0} fuses and privatizes $L_t$ under $l$.
  \item Line \ref{alg:su:h} reuses the repair function analysis of rolling update
    to find $h$ from $\loc{L_t}$.
  \item Line \ref{alg:su:apply} applies the repair function $h$ to the global nest $\glb{L_t}$.
\end{itemize}

\newcommand{\accxglb}[1]{\access{X_{#1,g}}{#1,g}{\bi}}
\newcommand{\accxloc}[1]{\access{X_{#1,l}}{#1,l}{\bi}}
\textbf{Applying the Repair Function $h$ (Split-K).}
After naive loop fusion and privatization,
the local nest $\loc{L_t}$ of the fusee reads from the local versions of its producing reductions,
which is incorrect.
Split-k repairs the global nest $\glb{L_t}$ and does not change $\loc{L_t}$,
keeping it parallelizable.
Another difference from rolling update is that split-k applies the repair function on the
right-hand side of the reducer.
The following \textit{rewrite pattern} shows how split-k update applies the repair function:
\begin{align*}
  \incorrect{\accxglb{t}}         & = \incorrect{\accxglb{t}} \cf \incorrect{\accxloc{t}} \\
  \Rightarrow \fixed{\accxglb{t}} & = \fixed{\accxglb{t}} \cf h\left(\accxloc{t},
  \accxloc{r}, \accxglb{r} \right)
  \numberthis \label{eq:su:rewrite}
\end{align*}
This equation shows a ``before $\Rightarrow$ after'' rewrite pattern.
The first line is the body of the global nest before the rewrite,
and the second line is after the rewrite.
$X_{t,g}$, $X_{t,l}$, $X_{r,g}$ and $X_{r,l}$ are the output tensors of
$\glb{L_t}$, $\loc{L_t}$, $\glb{L_r}$ and $\loc{L_r}$ respectively.
Tags are omitted because every tensor access refers to the final result of that memory location.
In the example (Figure \ref{fig:su-result}), split-k modifies the global loop nest \ssg{}
and the computation in the local \ssl{} is unchanged.
The global loop nest \ssg{} now has the store statement
\texttt{sum\_g[i] = sum\_g[i] + \uline{exp(max\_l[i, j0] - max\_g[i])} * sum\_l[i, j0]},
where the underlined term is the repair term.

\textbf{Correctness of Split-K Update.}
The definition of transformation correctness remains the same as rolling update:
split-k update produces a program where the reduce nest $\glb{L_t}$ produces
the same tensor values as the original loop nest $L_t$ before the fusion.

\begin{theorem}[Correctness of Split-K Update]
  \label{thm:su-correct}
  If split-k update successfully creates $\glb{L_t}$ and applies the repair in
  Eq.~\ref{eq:su:rewrite},
  then $X_{t,g}$ in the repaired program after the execution of $\glb{L_t}$
  equals $X_t$ in the original program after the execution of $L_t$.
\end{theorem}
We sketch a proof for this theorem here;
the full proof of this theorem is in Appendix \appref{sec:app:proofs}.

We first describe the values of the program before and after the transformation.
Before the transformation, the loop nest $L_t$ produces $X_t$.
The correctness proof of rolling update has shown the explicit expression for $X_t$
in Eq.~\ref{eq:original-explicit}.
We need to compare this expression of $X_t$ to the expression of $X_{t,g}$
produced by the global nest $\glb{L_t}$.
Because $\glb{L_t}$ reads from the local reduction results $X_{t,l}$,
we first write the expression for $X_{t,l}$ by writing down the body of $\loc{L_t}$ as a program:
\[
  (\texttt{for j}:) \qquad \accxloc{t} := \accxloc{t} \cf g\left(\accxloc{r}, \cij\right)
\]
We denote the number of iterations of the local reduction as $\loc{\bm}$,
\mbox{and express $\accxloc{t}$ explicitly:}
\[
  \incorrect{\accxloc{t}} = \foldbj{\loc{\bm}}{g\left(\accxloc{r}, \cij\right)}
\]
Combining $\incorrect{\accxloc{t}}$ and the body of $\loc{L_t}$ in Eq.~\ref{eq:su:rewrite},
we simplify the expression using the repair function $h$, which is shown in the full proof.
The simplified expression matches the before-transformation expression for $X_t$ in
Eq.~\ref{eq:original-explicit}.

\textbf{Tradeoffs of Split-K Update.}
Split-k update provides different tradeoffs from rolling update:
in addition to the benefits of operator fusion,
it offers more parallelism through the use of privatization
to create parallelizable local reductions.
On the other hand, it requires extra space for the local reduction results,
controlled by the tile size $k$.
Applying split-k update to attention produces kernels similar to
FlashDecoding~\cite{flash-decoding},
which is the state-of-the-art method for attention in decoding mode,
as decoding tolerates more space overhead and benefits greatly from more parallelism.

\subsection{Example Operators Amenable to \NAME{} Transformations}

\hspace{\parindent}\textbf{Attention.}
The attention operator is a sequence of four compute steps: matrix multiplication (matmul),
element-wise score computation, softmax, and another matmul.
Appendix \appref{sec:app:example-operators} provides a detailed description of the
attention computation.
\NAME{} fuses all the compute steps of attention into a single loop nest
by applying reduction fusion twice.
The first fusion applies to the two reductions in softmax,
similar to our motivating example in Figure \ref{fig:ru-example}.
The second fusion fuses the second matmul into the softmax.
Following the algebraic correction analysis (\S\ref{sec:repair-function}),
\NAME{} identifies the element-wise computation $g(r, c) = \exp(c - r)$
and reduction function $f(x, y) = x + y$.
Finally, Equation \ref{eq:h-solution} produces the repair term $h(t, r, r') = t \exp(r' - r)$.

\textbf{Performer.}
Performer~\cite{performers} is an alternative operator to attention
that approximates softmax with linear algebraic and element-wise computations.
Performer consists of many small compute steps and exhibits lower overall computational
complexity than attention.
\NAME{} fuses the many compute steps of performer into two loop nests,
applying reduction fusion four times.
The compute pattern $g$ in all cases and the $h$ repair term in one case are different
from attention.
Appendix \appref{sec:app:example-operators} describes these reduction fusion patterns in detail.

\textbf{Numerically stable $L^2$ norm.}
Computing the $L^2$ norm of a vector
by definition can overflow when the squared values are too large.
It is more numerically stable to scale the entries in the vector first:
$||\boldsymbol{x}||_2 = m \sqrt{\sum_i(x_i/m)^2}$ where $m = \max_i |x_i|$.
This numerically stable algorithm exists in widely used linear algebra libraries such as
LAPACK~\cite{lapack99}.
\NAME{} fuses the second reduction (summation) into the first (maximum) so that
the $L^2$ norm is computed in a single pass.
Following the algebraic correction analysis (\S\ref{sec:repair-function}),
it identifies the reduction function $f(x, y) = x + y$
and the element-wise computation $g(r, c) = (c/r)^2$.
Using Equation \ref{eq:h-solution}, it finds the repair function $h(t, r, r') = t (r^2/r'^2)$.

\textbf{Dynamic scaling quantization on DNN operators.}
Low-precision training and inference often computes statistics (e.g. minimum and maximum)
over the input and output tensors to derive scaling and quantize the tensors.
\NAME{} can fuse many variants of this computation with the surrounding layers.
One such example is computing RMSNorm: $y_i = x_i / \sqrt{s + \varepsilon}$ where $s =
\mathrm{mean}(x^2)$, and then finding the per-row max of the output: $m = \max_i y_i$.
The mean and max computations are two reductions that \NAME{} fuses.
\NAME{}'s algebraic correction analysis identifies:
\vspace{-0.5em}
\[
  g(r, c) = \frac{c}{\sqrt{r + \varepsilon}}, \quad f(x, y) = \max(x, y), \quad \text{and} \quad
  h(t, r, r') = \frac{\sqrt{r + \varepsilon}}{\sqrt{r' + \varepsilon}} t
  \vspace{-0.5em}
\]
Notably, \NAME{} validates that $h$ distributes with $\max$ (condition \ref{eq:h-conditions-2}),
because $\sqrt{r + \varepsilon} / \sqrt{r' + \varepsilon}$ is always non-negative,
and $h$ monotonically increases with $t$.
}

\section{Translating from Loop-scalar IR to Tile IR}
\label{sec:l2t}
{\def\ttt{\texttt}
\def\taccess{id\ttt{[}expr\wild{}\ttt{]}}
\def\as{\ :=\ }
\def\wild{\!*\!}

\begin{figure}[tb]
  \includegraphics[width=\textwidth,trim=0mm 24mm 50mm
  1mm,clip,page=2]{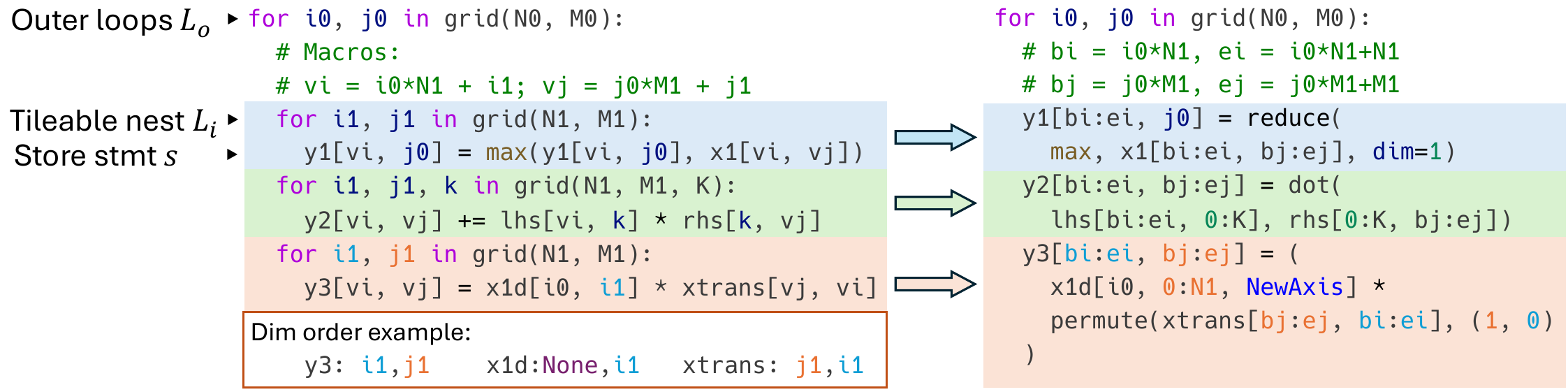}
  \caption{An example translation from a loop-scalar program (left) to a tile program (right),
    with three tileable loop nests to translate.
    The bottom of the LHS program shows an example of dimension order reconciliation
    for the last loop nest.
  The LHS program uses shorthands \texttt{vi} and \texttt{vj} to express loop-tiled indices.}
  \label{fig:example-l2t}
\end{figure}

Tile-based compilers such as Triton~\cite{triton} and Pallas~\cite{pallas}
excel in low-level optimizations to generate efficient device code optimized for local
hardware features,
such as tensor cores for matrix multiplication.
To take advantage of tile compilers, \NAME{} translates loop-level programs
operating on scalar elements
to tile-level programs in a process known as \textit{tensorization}.
While tensorization has been studied extensively~\cite{tensorize-amos,
tensorize-volta-tc, tensorize-unit},
\NAME{} adds a novel \textit{dimension reconciliation} step
that generalizes the translation to a larger class of programs.
\NAME{} translates loop programs to \textit{HTile} IR, a novel, custom tile IR
that we design to resemble existing tile languages,
and generates code from HTile IR to different tile languages such as Triton.

Figures \ref{fig:loop-ir-syntax} and \ref{fig:tile-ir-syntax} show the abstract syntax
of \NAME{}'s loop-scalar IR and HTile IR, respectively.
Compared to the loop-scalar IR, HTile IR has the same loop and sequential statements,
and replaces scalar expressions and store statements with tensor tile expressions and
tile store statements.
HTile IR also adds tile-specific operations like dimension operations (broadcasting and
permutation), reductions, and dot products.
These operations are selected because they receive wide support from tile compilers,
which is an effect of the hardware features available on today's GPUs.

\begin{wrapfigure}{r}{0.55\columnwidth}
  \begin{algorithm}[H]
    \small
    \newcommand{\func}[1]{\textup{\textsc{#1}}}
    \newcommand{\highlight}[1]{\textcolor{OliveGreen}{#1}}
    \SetInd{0.3em}{0.6em}
    \SetKwFor{Match}{match}{}{}
    \SetKwFor{Case}{case}{:}{}
    \SetKw{Continue}{continue}
    \caption{\func{TranslateLoopToTile}($s_0$)}
    \label{alg:l2t}
    \KwIn{$s_0$: the statement to translate, in loop-scalar IR.}
    \KwOut{The translated statement in HTile IR.}
    \For{$s \in \func{VisitScalarStoreStmts}(s_0)$}{ \label{alg:l2t:foreach-store}
      $(L_o, L_i)$ := \func{TraverseUpForTileableNest}($s_0$, $s$)\; \label{alg:l2t:find-nest}
      \For{$e \in \func{VisitTensorAccessExprs}(s)$}{ \label{alg:l2t:for-expr-begin}
        $e_t$ := \func{CreateTensorTileAccess}($e$, $L_i$)\; \label{alg:l2t:create-tile}
        $d$ := \highlight{\func{ExtractDimOrder}($e$, $L_i$)}\; \label{alg:l2t:ext-order}
        $e_t'$ := \highlight{\func{ReconcileDimOrderToLoopOrder}($e_t$, $d$, $L_i$)}\;
        \label{alg:l2t:recon}
        \func{Replace}(in=$s$, from=$e$, to=$e_t'$)\;
      }
      \label{alg:l2t:for-expr-end}
      $s'$ := \func{MatchComputePattern}($s$)\; \label{alg:l2t:match-pattern}
      \func{Replace}($s_0$, $L_0$, $s'$)\;
    }
    \Return{$s_0$}
  \end{algorithm}
\end{wrapfigure}

Figure \ref{fig:example-l2t} shows a before-after example of a translated tile program.
Algorithm \ref{alg:l2t} outlines the translation procedure with three main steps:
tileable nest detection (algorithm lines \ref{alg:l2t:foreach-store}-\ref{alg:l2t:find-nest}),
access region division (lines \ref{alg:l2t:for-expr-begin}-\ref{alg:l2t:create-tile}),
and dimension order reconciliation (lines \ref{alg:l2t:ext-order}-\ref{alg:l2t:recon} in green).
Dimension order reconciliation makes the translation more general than typical
tensorization algorithms
that rely on pattern matching to match tile computation to specific hardware features
(e.g., detecting transposed input for TensorCore matmuls).

\textbf{Tileable Nest Detection} finds the tileable loop nest (line \ref{alg:l2t:find-nest})
for each store statement $s$ by walking up the AST from $s$,
stopping at the first dynamically sized loop or non-loop statement.
This take-until procedure produces the tileable loop nest $L_i$ of $s$.
Then it continues to collect any outer loops enclosing $L_i$ and returns them as $L_o$.
For example, Figure \ref{fig:example-l2t} contains three tileable loop nests,
the first being $L_i = (\texttt{i1}, \texttt{j1})$ (with blue background),
and $L_o = (\texttt{i0}, \texttt{j0})$ are the outer loops for all three nests.
Next, $L_i$ will be translated into a single tile store statement.

\textbf{Access Region Division} (line \ref{alg:l2t:create-tile})
replaces each tensor access $\taccess{}$ in $s$ with a tensor \textit{tile} access.
It converts \texttt{x1d[i0, i1]}, an access in the third nest of Figure \ref{fig:example-l2t},
into \texttt{x1d[i0:i0+1, 0:N1]}.
Each index is a \textit{slice} with a start and an end index.
\NAME{} uses existing tensorization algorithms (built on TVM~\cite{tvm})
to implement this step, finding regions of access under loop nests.

\begin{figure}[t]
  \centering
  \begin{minipage}[t]{.5\textwidth}
    \vspace{0pt}\small\centering
    \begin{align*}
      stmt        & \as loop \mid store \mid seq\_stmt                                      \\
      seq\_stmt   & \as stmt\ stmt+                                                         \\
      iter\_space & \as \ttt{range}(expr) \mid \ttt{grid}(expr+)                            \\
      loop        & \as \ttt{for } id+\!\ttt{ in } iter\_space\ttt{: } stmt                 \\
      store       & \as \taccess{} = expr                                                   \\
      expr        & \as lit \mid id \mid \taccess \mid op\ttt{(}expr\wild{}\ttt{)}          \\
      op          & \as + \mid - \mid \ttt{exp} \mid \ttt{log} \mid \ttt{select} \mid \dots
    \end{align*}
    \captionof{figure}{\NAME{} loop-scalar IR syntax.}
    \label{fig:loop-ir-syntax}
  \end{minipage}
  \hfill
  \begin{minipage}[t]{.49\textwidth}
    \vspace{0pt}\small\centering
    \begin{align*}
      stmt    & \as loop \mid store_t \mid stmt*                              \\
      idx     & \as expr \mid expr\ttt{:}expr \mid \ttt{NewAxis}              \\
      store_t & \as id\ttt{[}idx\wild\ttt{]} = expr_t                         \\
      expr_t  & \as id\ttt{[}idx\wild\ttt{]} \mid op\ttt{(}expr_t\wild\ttt{)} \\
      & \mid \ttt{permute(}expr_t\ttt{, order=(}i\wild\ttt{))}        \\
      & \mid \ttt{reduce(}op\ttt{, }expr_t\ttt{, dim=}i\ttt{)}        \\
      & \mid \ttt{dot(}expr_t\ttt{, }expr_t\ttt{, acc=}expr_t\ttt{)}
    \end{align*}
    \captionof{figure}{\NAME{} HTile IR syntax.
    Non-terminals that are the same as in Figure~\ref{fig:loop-ir-syntax} are omitted.}
    \label{fig:tile-ir-syntax}
  \end{minipage}
\end{figure}

\subsection{Dimension Order Reconciliation}
\label{sec:dim-recon}

The previous steps extract tile regions from store statements loop nests in the loop-scalar IR.
A loop nest holds more information, however:
the order of loops also naturally encodes how the dimensions of tensors are traversed.
The third loop nest in Figure \ref{fig:example-l2t} highlights what is missing from just
tile regions.
Simply assembling the tiles from the Access Region Division step would produce
\texttt{y3[bi:ei, bj:ej] = x1d[i0:i0+1, 0:N1] * xtrans[bj:ej, bi:ei]},
which is incorrect and leaves out the broadcast and permutation operations
that the original loop nest expressed.

\definecolor{icolor}{HTML}{0F9ED5}
\definecolor{jcolor}{HTML}{E97132}
\definecolor{ncolor}{HTML}{78206E}
\def\i{\textcolor{icolor}{i1}}
\def\itext#1{\textcolor{icolor}{#1}}
\def\j{\textcolor{jcolor}{j1}}
\def\jtext#1{\textcolor{jcolor}{#1}}
\def\n{\textcolor{ncolor}{None}}
\def\ntext#1{\textcolor{ncolor}{#1}}
To preserve order information, we take inspiration from Einstein notation and
\textit{einops} expressions~\cite{einops}, which assigns a symbol per dimension
and encodes the dimension orders in a string like \texttt{i,ji -> ij}.
We use the loop variables in $L_i$ as the set of symbols,
and extract a dimension order $d$ for each tensor access $e$.
Examples on the bottom left of Figure \ref{fig:example-l2t} show:
\texttt{y3[\itext{vi}, \jtext{vj}]} yields order \texttt{(\i,\j)},
\texttt{xtrans[\jtext{bj}, \itext{bi}]} yields \texttt{(\j,\i)},
and \texttt{x1d[\ntext{i0}, \itext{i1}]} yields \texttt{(\n,\i)}.
\texttt{None} is a placeholder for an index that does not depend on the iteration variables
from loops in $L_i$.

Next, we match $d$ to the order of loop variables in $L_i$, denoted as $d_L$.
As we rearrange $d$, we apply a corresponding operation on the tile access $e$ to obtain
correct access $e_t$. For example, to bring tensor access $e =
\ttt{xtrans[\jtext{bj}:\jtext{ej}, \itext{bi}:\itext{ei}]}$ (with $d = \texttt{(\j,\i)}$)
to match loop order \texttt{(\i,\j)}, we apply a matching transpose $e_t =
\ttt{permute(xtrans[\jtext{bj}:\jtext{ej}, \itext{bi}:\itext{ei}], (1, 0))}$.
Any $d$ can be rearranged to match $d_L$ in three steps:
\begin{itemize}
  \item \textbf{Dimension squeezing} removes a \texttt{None} from $d$
    and converts the slice on that dimension in $e_t$ to a scalar index.
    The slice is guaranteed to have an extent of $1$, because it does not use loops from $L_i$.
    For $d = \texttt{(\n, \i)}$ and $e_t = \texttt{x1d[i0:i0+1, 0:N1]}$,
    we discard the \texttt{\ntext{None}} from $d$ and convert $e_t$ to \texttt{x1d[i0, 0:N1]}.
  \item \textbf{Permutation} arg-sorts $d$ to $d_L$ and produces a permutation order $\sigma$.
    Sorting is possible because we have removed \texttt{None} from $d$, and $d \subseteq d_L$.
    For $d = \texttt{(\j, \i)}$, we arg-sort $d$ to get $\sigma = (1, 0)$
    and apply $\texttt{permute(}e_t\texttt{, order=(1, 0))}$ to $e_t$.
  \item \textbf{Dimension unsqueezing} (dimension expansion)
    inserts a missing variable from $d_L$ into $d$,
    and adds a \texttt{NewAxis} index on $e_t$, which creates a size-1 dimension.
    For $d = \texttt{(\i,)}$ and $e_t = \texttt{x1d[i0, 0:N1]}$,
    \texttt{\j} is missing at the end of $d$, so we insert a \texttt{NewAxis} to $e_t$
    to get \texttt{x1d[i0, 0:N1, NewAxis]},
    which brings the tile to 2-D of size $N_1 \times 1$.
\end{itemize}

On lines \ref{alg:l2t:ext-order}-\ref{alg:l2t:for-expr-end} in the algorithm,
we repeat dimension reconciliation for each tensor access $e$ in $s$,
which produces a tile store statement that is faithful to the original loop nest.
Lastly, the algorithm matches this tile store to the types of tile computation that HTile IR allows
(dot product, reduction, element-wise; line \ref{alg:l2t:match-pattern}),
based on the number of reduction loops in the loop nest,
and replaces the inner loop nest $L_0$ with this tile store $s'$.

\subsection{Generality of the Translation Algorithm}

The translation algorithm supports many programs representable in the loop-scalar IR.
As HTile IR is designed to fit the capabilities of \NAME{}'s tile optimizer,
it rejects a program if any tensor access uses an inner loop variable (from $L_i$) multiple times,
such as \texttt{tensor[i1, i1]},
or uses any inner loop variable non-linearly, such as \texttt{tensor[i1 * i2]}.
These usages are not supported by the tile optimizer and are rare in tensor programs used
in machine learning.
}

\section{Implementation}
\label{sec:impl}
{\NAME{} is built on top of TVM and Triton tensor compilers.
We implement \NAME{}'s loop-scalar IR and HTile IR as extensions of TVM TensorIR.
\NAME{}'s two inputs, tensor expression and schedule template,
are based on TVM's tensor expression (TE) and scheduling language respectively.
\NAME{} extends TVM's scheduling language by adding transformation primitives
such as rolling update.
\NAME{} uses Triton as its tile optimizer and code generator,
and \NAME{} has a translator from HTile IR to Triton's Python DSL.
We use SymPy, a Python library for solving symbolic equations,
to implement the analysis for the repair function $h$ in rolling update and split-k update.

\NAME{} is implemented in 10K lines of C++ and Python code,
with 3K lines for loop fusion transformations
and 1.3K lines for loop-to-tile translation.
\NAME{} adapts TVM MetaSchedule~\cite{metaschedule} as its autotuner, reusing its search algorithm.
We register our transformations with MetaSchedule
so that it knows how to work with \NAME{} schedules.
MetaSchedule is a cost model-guided autotuner that repeats the following steps:
produce 1024 schedules, predict their performance,
and run 16 of them on the target hardware (called \textbf{\textit{empirical measurements}}).
Every prediction invokes the template-guided optimizer in \NAME{},
while every empirical evaluation fully generates the program,
applying our loop-to-tile translation and Triton-based tile optimization.
}

\section{Experimental Methodology}
\label{sec:method}
{\hspace{\parindent}\textbf{GPU platforms.}
We choose two
\textit{datacenter devices} Nvidia A100~\cite{nvidia-a100} (SXM4 40 GB version) and AMD MI300;
and two \emph{desktop devices} Nvidia RTX A5000~\cite{nvidia-a5000} and Nvidia RTX 6000
Ada~\cite{ nvidia-6000}.

\begin{table}[b]
  \small
  \centering\vspace{0.02in}
  \caption{LLM operators used in our evaluation. PF is short for prefill and DC for
  single-token decoding.}
  \label{tab:operators}
  \begin{tabular}{l l l} \Xhline{2\arrayrulewidth}
    \textbf{Operator} & \textbf{Description}                & \textbf{Base Arch.}   \\ \hline
    Global (PF)       & Global (plain) attention            & ViT-L/16              \\
    Causal (PF/DC)    & Attn + causal mask                  & GPT3 6.7B             \\
    ALiBI (PF/DC)     & Attn + ALiBi bias + causal mask     & MPT 7B                \\
    GQA (PF/DC)       & Attn + GQA heads  + causal mask     & LLama3 70B            \\
    SoftCap (PF/DC)\qquad\    & Attn + SoftCap bias + GQA heads  + causal mask \qquad\  &
    {Gemma2 27B} \\
    Window (PF)       & Attn + windowed mask                & {Gemma2 27B}                       \\
    \Xhline{2\arrayrulewidth}
  \end{tabular}
\end{table}

\textbf{Attention-based Operator Experiments.}
We evaluate \NAME{} on 10 attention-based operators shown in Table~\ref{tab:operators}.
Each operator is extracted from a specific large language model (LLM)
or vision language model (VLM) architecture, shown on the \textbf{Base Arch.} column.
We profile operators in prefill (PF) mode:
where the query sequence length $s_q$ equals key/value length $s_{kv}$,
and decoding (DC) mode, where $s_q = 1$.
Variations on masking (Global, Causal, and Window) are indistinguishable in decoding,
so we select the Causal variant out of the three.
All operators run in inference mode at float16 (half) precision.
We profile these operators at varying sequence lengths:
2\verb+^+7 = 128, 2\verb+^+8 = 256, ..., 2\verb+^+15 = 32768,
and a batch size of 1 unless otherwise specified.
We refer to the triple of operator, input shape, and GPU as a \textbf{setup},
and we evaluate $10 \times 8 \times 4 = 320$ setups.
We run \NAME{}'s autotuner for 128 empirical measurements per setup.

\textbf{Non-Attention Operator Experiments.}
We evaluate \NAME{} on 3 non-attention operators:
numerically stable $L^2$ norm, RMSNorm with dynamic scaling quantization, and performer,
profiled on Nvidia RTX 6000 Ada.
For performer, we use the same base architecture as defined in the performer official
implementation \cite{performer-pytorch-repo}: 8 attention heads, per-head dimension 64,
and number of features 256.
We maintain a batch size of 1 and vary the sequence length $L$ from 512 to 32768.
For $L^2$ norm and RMSNorm, we direct their reduction along the rows,
and use a 2 dimensional input of shape $N \times M$.
$N$ varies from 32 to 256, while $M$ varies from $2^{10} = 1024$ to $2^{17} = 131072$.

\textbf{Baselines.}
We list our 10 baselines and underline the names we refer to them by in evaluation.
Our baselines include 4 tensor compilers:
Triton 3.2, FlexAttention (\uline{FlexAttn}) 2.6.0, \uline{Mirage} 0.2.2, and \uline{TVM} 0.18.
Triton requires a user to write the kernel in its frontend language,
so we compare three mainstream implementations in Triton:
\uline{OpenAI Triton} \cite{openai-fused-attn}, \uline{Tri-Dao Triton} \cite{flash-attn-repo},
and \uline{XFormers Triton} \cite{xformers}.
Our baselines also include 4 optimized inference libraries:
\uline{PyTorch} 2.6.0 \cite{pytorch}, \uline{cuDNN} 9.11.0,
\uline{Tri-Dao Cutlass} 2.7.4 (Dao-AILab's implementation) \cite{flash-attn-repo},
and \uline{FlashInfer} 0.2.4 \cite{flashinfer}.
Each implementation supports a subset of the setups we evaluate.
For PyTorch, we use \verb|torch.compile(| \verb|backend='inductor')| to ensure high performance.
We evaluate Mirage on A100, on the Global operator
over a grid of $s_q \neq 1$ and $s_{kv}$, which is consistent with the evaluation in the
Mirage paper.
For non-attention operators, we compare \NAME{} kernels against PyTorch kernels optimized using
\verb|torch.compile(backend='inductor')|.

\textbf{Profiling and Performance Reporting.}
For any kernel we profile, we run the kernel once to warm up and discard the result,
and run 15 times to report the mean latency.
We profile the kernels with Nsight Systems (\texttt{nsys}) on Nvidia GPUs
and ROCprofiler (\texttt{rocprofv3}) on AMD GPU.
Both profilers trace the program and measure the latency of every kernel in the program.
We report \textbf{speedup} of \NAME{} over the best baseline:
if \NAME{}'s kernel has mean latency $t_0$ and the baselines have $t_1, \dots t_n$,
then our speedup is $\min(t_1, \dots, t_n) / t_0$.
We also report \textbf{relative performance} of a baseline relative to \NAME{},
which is $t_0 / t_i$ for the $i$-th baseline.
To average these metrics across setups, we compute \mbox{geometric means (geomeans).}
}

\section{Evaluation}
\label{sec:eval}
{\def\cross{\ding{55}}

\subsection{\NAME{} vs.~Tensor Compilers}
\label{sec:eval-vs-vendor}

\begin{wraptable}{r}{0.48\columnwidth}
  \centering
  \small\vspace{-.07in}
  \caption{Speedup of \NAME{} relative to the best compiler baseline, for all 10
  operators on 4 GPUs.}
  \label{tab:rq1}
  \def\dgbox{\diagbox[width=7.5em, height=2.2em]{Operator}{GPU}}
  \setlength{\tabcolsep}{3pt}
  \renewcommand{\arraystretch}{0.93}
  \begin{tabular}{l|cccc} \toprule
    \dgbox           & 6000Ada & A5000 & A100 & MI300 \\ \midrule
    Global (PF)      & 1.03    & 1.05  & 1.10 & 1.84  \\
    Causal (PF)      & 1.06    & 1.06  & 1.05 & 1.29  \\
    GQA (PF)         & 1.02    & 1.05  & 1.06 & 1.32  \\
    Causal (DC)      & 1.06    & 1.08  & 1.11 & 3.26  \\
    GQA (DC)         & 1.45    & 1.37  & 2.21 & 1.55  \\
    ALiBi (PF)       & 1.56    & 1.61  & 1.71 & 2.02  \\
    ALiBi (DC)       & 1.28    & 1.27  & 2.65 & 3.32  \\
    SoftCap (PF)     & 1.08    & 1.07  & 1.03 & 1.42  \\
    SoftCap (DC)     & 1.08    & 0.95  & 1.86 & 2.55  \\
    Window (PF)      & 1.03    & 1.02  & 0.99 & 1.23  \\
    \bottomrule
    \textbf{Average} & 1.15    & 1.14  & 1.38 & 1.85  \\
    \bottomrule
  \end{tabular}
\end{wraptable}
\textbf{Overall Trends.}
Table \ref{tab:rq1} presents the average speedup of \NAME{} over the best compiler baseline
where each cell is a geomean over the input sequence lengths.
Compilers are a broad term; here we only include compilers that raise the
abstraction level above hardware vendor languages, such as tile compilers. See
\S\ref{sec:rw} for a detailed explanation.
Out of 320 setups we evaluated, \NAME{} achieves better or equal performance
compared to all other compilers on 284 setups.
\NAME{} shows an improvement over the baselines for all four GPU architectures,
ranging from $1.15\times$ to $1.85\times$ and
indicating the portability of our approach across GPU architectures.
Across all setups,
\NAME{} achieves $1.35\times$ the performance of the best compiler baseline.
All kernels for PF benchmarks use Rolling Update,
and all kernels for DC \mbox{benchmarks use Split-K Update.}

\noindent\textbf{Detailed Results for Setups}.
Figure~\ref{fig:rq1-first5-ops} presents detailed performance for every setup for five
operators, chosen for prevalence in existing LLMs.
Each plot shows relative performance (y-axis) of other compilers normalized to \NAME{}
for \mbox{different sequence lengths (x-axis).}

On Nvidia GPUs, \NAME{} achieves the best performance in a majority of setups.
TVM performance is low as its fusion legality check
forbids reduction fusion, resulting in multiple kernels with increased memory transfer.
Most tile-based implementations show better performance than TVM
and improve their performance as the sequence length increases.
For prefill (PF) operators, \NAME{} achieves higher performance for shorter sequence lengths.
\NAME{} and FlexAttn benefit from autotuning
and have higher performance than other baselines that rely {on Triton heuristics.}

\begin{figure*}[t]
  \centering\vspace{-.03in}
  \begin{subfigure}[b]{0.95\textwidth}
    \includegraphics[width=\textwidth]{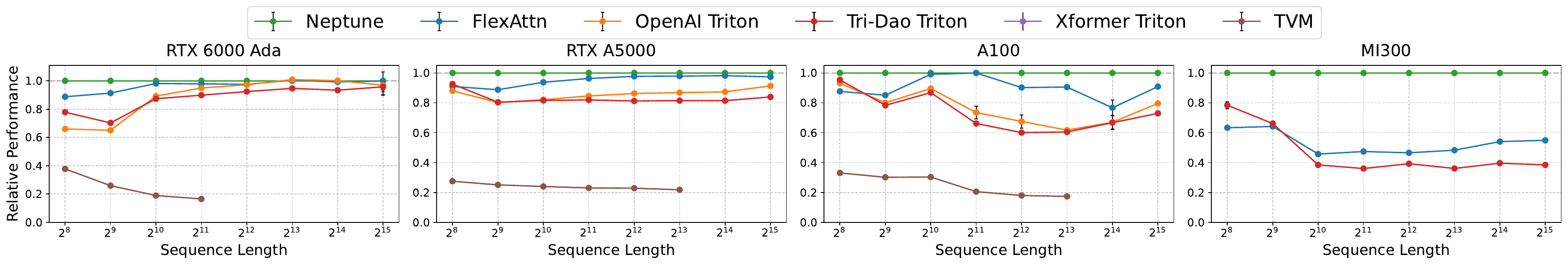}
    \vspace{-1.7em}
    \caption{Global (PF) operator (uses rolling update; higher is better).}
    \label{fig:rq1-first-op}
  \end{subfigure}
  \begin{subfigure}[b]{0.95\textwidth}
    \includegraphics[width=\textwidth]{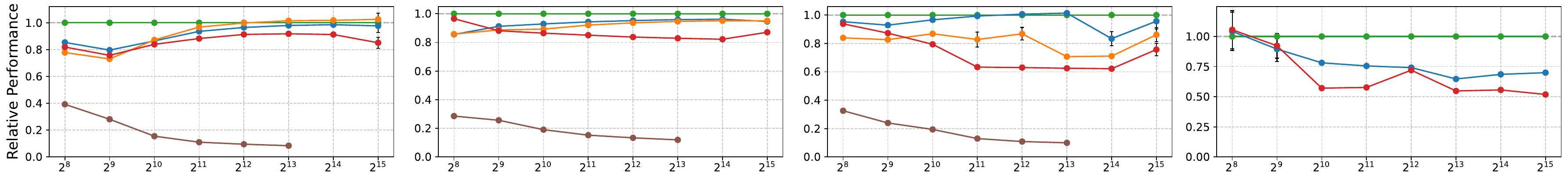}
    \vspace{-1.7em}
    \caption{Causal (PF) operator (uses rolling update; higher is better).}
  \end{subfigure}
  \begin{subfigure}[b]{0.95\textwidth}
    \includegraphics[width=\textwidth]{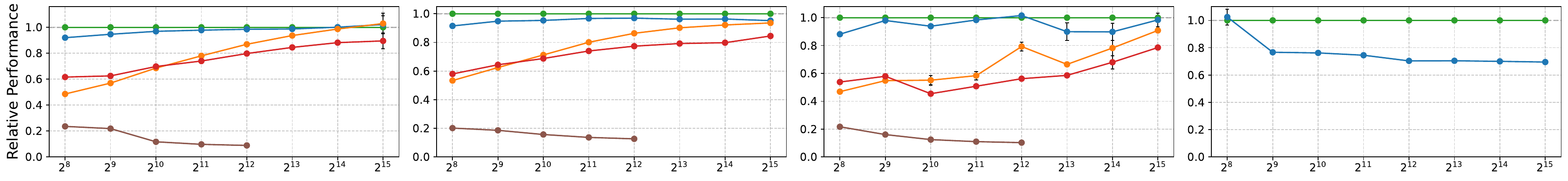}
    \vspace{-1.7em}
    \caption{GQA (PF) operator (uses rolling update; higher is better).}
  \end{subfigure}
  \begin{subfigure}[b]{0.95\textwidth}
    \includegraphics[width=\textwidth]{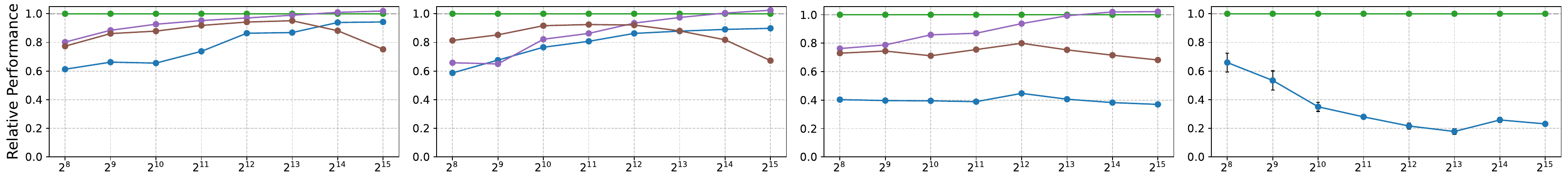}
    \vspace{-1.7em}
    \caption{Causal (DC) operator (uses split-k update; higher is better).}
  \end{subfigure}
  \begin{subfigure}[b]{0.95\textwidth}
    \includegraphics[width=\textwidth]{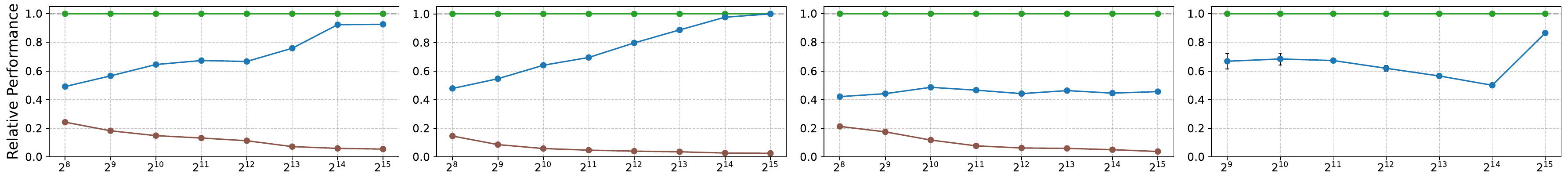}
    \vspace{-1.7em}
    \caption{GQA (DC) operator (uses split-k update; higher is better).}
    \label{fig:rq1-op5}
  \end{subfigure}
  \caption{Performance of \NAME{} kernels vs.~kernels by other tensor compilers.
    Each y-axis shows relative performance of all compilers normalized to \NAME{}. The
  plots for the remaining benchmarks are in Appendix~\appref{sec:app:add-eval}.}
  \label{fig:rq1-first5-ops}
  \vspace{-.05in}
\end{figure*}

For decoding (DC) operators, we evaluate XFormer Triton since the other two Triton
kernels do not support them.
TVM provides higher performance for Causal (DC) than for other operators,
because Causal (DC) is a matrix-vector multiplication operator
where fusion has limited effect on its memory-boundness.
\NAME{} delivers consistently better performance on GQA (DC) because
it uses masks to apply TensorCore to the dot product in the computation
that would otherwise have too few input matrix rows for TensorCore.
There is no single baseline that provides high performance like \NAME{} does
across all operators~and~GPUs.

Only FlexAttn and Tri-Dao Triton (limited to Global and Causal PF) support AMD GPU.
\NAME{} delivers consistently high performance compared to available baselines.

\textbf{Programmability.}
Triton-based baselines require a user to write the kernel in their DSL,
and FlexAttn uses multiple attention-specific templates developed by the framework authors.
For instance, the Tri-Dao Triton kernel is 650 lines of code,
while \NAME{}'s input is vanilla attention (38 lines) and schedule (28 lines) in total;
see Appendix~\appref{sec:app:neptune-inputs} \mbox{for these inputs.}

\begin{table}[t]
  \def\sqkv{\diagbox[height=1.8em]{$s_q$}{$s_{kv}$}}
  \begin{minipage}[t]{0.43\linewidth}
    \centering\small
    \renewcommand{\arraystretch}{0.9}
    \caption{Speedup of \NAME{} relative to Mirage on global attention,
      over a grid of $s_q$ and $s_{kv}$ where $s_q \leq s_{kv}$.
      Cells where $s_q > s_{kv}$ has an em-dash (---).
    Cells where Mirage fails to find a valid kernel has a cross (\cross). The GPU is A100.}
    \label{tab:rq1-mirage}
    \vspace{.04in}
    \begin{tabular}{l|ccccc} \Xhline{2\arrayrulewidth}
      \sqkv & 128    & 256  & 512  & 1024 & 2048   \\ \hline
      16    & 1.98   & 2.25 & 1.60 & 1.30 & 4.10   \\
      32    & 2.10   & 1.44 & 1.47 & 1.25 & 1.39   \\
      64    & 1.63   & 1.92 & 1.32 & 1.29 & 1.58   \\
      128   & \cross & 1.83 & 2.40 & 1.63 & 1.85   \\
      256   & ---    & 2.04 & 3.35 & 4.40 & 6.40   \\
      512   & ---    & ---  & 3.64 & 6.01 & \cross \\
      1024  & ---    & ---  & ---  & 6.71 & \cross \\
      2048  & ---    & ---  & ---  & ---  & \cross \\
      \Xhline{2\arrayrulewidth}
    \end{tabular}
  \end{minipage}
  \hspace{0.03\linewidth}
  \begin{minipage}[t]{0.52\linewidth}
    \centering\small
    \caption{Speedup of \NAME{} relative to the best manually optimized library baseline,
    for 8 operators on 4 GPUs. Each cell is an average over input sequence lengths.}
    \label{fig:rq2}
    \def\dgbox{\diagbox[width=7.5em, height=2.2em]{Operator}{GPU}}
    \renewcommand{\arraystretch}{0.90}
    \begin{tabular}{l|ccccc} \toprule
      \dgbox           & 6000Ada & A5000 & A100 & MI300 \\ \midrule
      Global (PF)      & 0.96    & 0.95  & 0.84 & 0.93  \\
      Causal (PF)      & 0.97    & 0.89  & 0.81 & 0.68  \\
      GQA (PF)         & 0.93    & 0.85  & 0.80 & 0.64  \\
      Causal (DC)      & 0.99    & 0.98  & 0.99 & 5.32  \\
      GQA (DC)         & 1.09    & 1.17  & 1.24 & 2.14  \\
      ALiBi (PF)       & 1.65    & 1.36  & 1.24 & 0.98  \\
      ALiBi (DC)       & 1.07    & 1.12  & 1.17 & 5.71  \\
      Window (PF)      & 0.85    & 0.67  & 0.70 & 0.46  \\
      \bottomrule
      \textbf{Average} & 1.04    & 0.98  & 0.95 & 1.36  \\
      \bottomrule
    \end{tabular}
  \end{minipage}
  \vspace{0.02in}
\end{table}

\textbf{Mirage Superoptimizer.}
We discuss Mirage separately, as Mirage does not consistently find valid kernels for all
input shapes
and only supports variants of Global and Causal operators.
Table \ref{tab:rq1-mirage} shows the speedup of \NAME{} relative to Mirage.
The columns and rows vary by $s_{kv}$ and $s_q$ respectively, where $s_q \leq s_{kv}$.
Cells where $s_q > s_{kv}$ are filled with an em dash (---),
while cells where Mirage fails to find a valid kernel have a cross (\cross).
The results show $2.21\times$ lower latency (geomean) across all tested shapes,
while \NAME{} provides deterministic correctness guarantees (unlike Mirage's probabilistic),
and we observed better numerical accuracy with \NAME{}.

\vspace{-.05in}
\subsection{\NAME{} vs.~Manually Optimized Libraries}

Table \ref{fig:rq2} shows the performance of \NAME{} kernels vs.~kernels
from manually optimized libraries.
The y-axis shows average relative performance:
each framework's performance relative to \NAME{}, averaged over sequence lengths.
The error bar depicts the range of relative performance over sequence lengths:
if a framework does better on some shapes and worse on others, it will have a large error bar.
Table \ref{fig:rq2} omits the SoftCap operator (both prefill and decoding),
because existing baseline libraries do not support it.
Therefore, this table shows $4 \times 8 \times 8 = 256$ setups.

Out of 256 setups, \NAME{} has better or equal performance
compared to all libraries on 101 setups.
On average of all setups, \NAME{} delivers geomean $1.07\times$
the performance of the best library baseline.
There is not a single library that consistently delivers the best performance
or outperforms \NAME{} on all setups.
For example, cuDNN has the highest performance on most short-sequence setups,
and low performance for long sequences.
In contrast, CUTLASS implementations have higher performance on longer sequences
for many operators, such as Global (PF) and GQA (DC).
Appendix~\appref{sec:app:add-eval} Figure \ref{fig:appendix-rq2-all-ops} shows these trends
over all input shapes.

The high performance of kernel libraries is a result of manual optimization:
they show better performance on the more popular operators
(Global, Causal and GQA PF) and GPUs (e.g.~A100),
and lower performance than \NAME{} otherwise.
\NAME{} failed to optimize Window (PF) because
its TVM-based loop analysis could not identify the proper condition for loop partitioning.

\vspace{-.05in}
\subsection{Scalability Test}

Optimizations in \NAME{} involve various trade-offs that introduce a small amount of
resource overhead,
such as register and shared memory (SMEM) usage, for significant performance gains.
To show that \NAME{}'s resource usage is acceptable, we stress-test \NAME{} on increasing
batch sizes
until we reach the limit of available GPU memory.
Figure \ref{fig:throughput} shows the throughput (in TFLOPs/sec) of \NAME{}'s kernel
and multiple baselines for the GQA (PF) operator on RTX A5000.
Appendix~\appref{sec:app:add-eval} presents throughput results over all operators and GPUs.
We use a sequence length of 8192 as it allows us to test more batch sizes up to 32.
The throughput of all kernels decreases as input size increases,
as longer workloads are more likely to trigger GPUs' clock and power throttling.
\NAME{}'s kernel remains close to the best baseline (slightly behind cuDNN)
\mbox{for all tested batch sizes.}

\begin{figure}[t]
  \begin{minipage}[t]{0.48\columnwidth}
    \centering
    \includegraphics[width=\columnwidth]{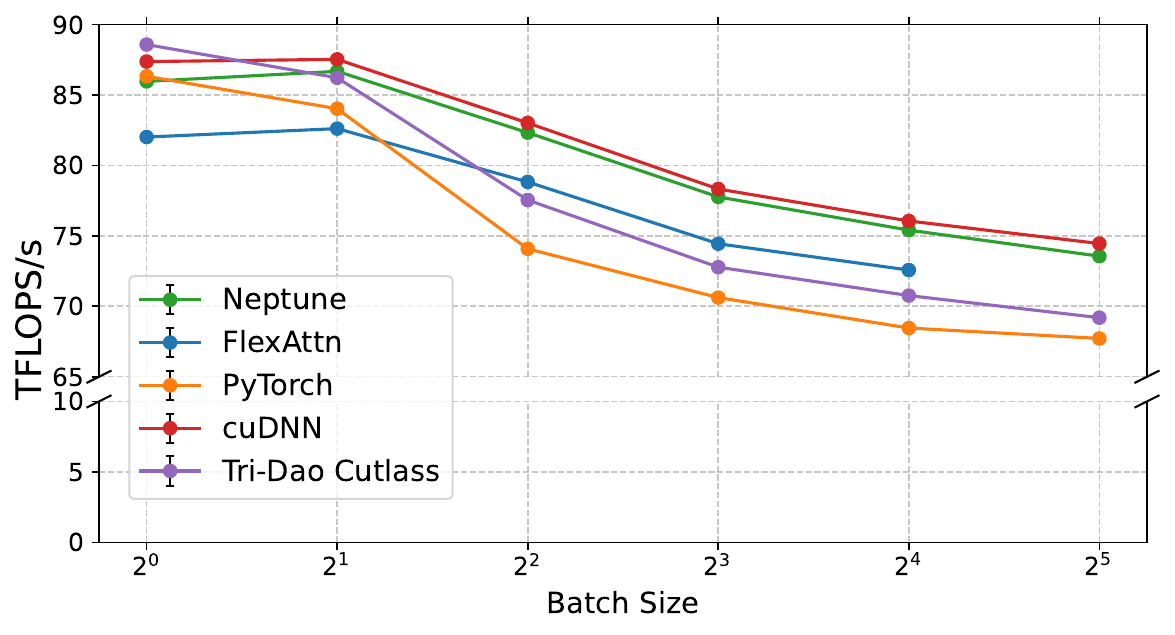}
    \caption{
      Throughput of \NAME{} kernels and baselines over increasing batch sizes,
      evaluated on RTX A5000 for the GQA (PF) operator.
    The y-axis shows throughput in TFLOPS/sec, and the x-axis shows batch size.}
    \label{fig:throughput}
  \end{minipage} \hfill
  \begin{minipage}[t]{0.50\columnwidth}
    \centering
    \includegraphics[width=\columnwidth, clip, trim=0 2.5mm 0 2mm]{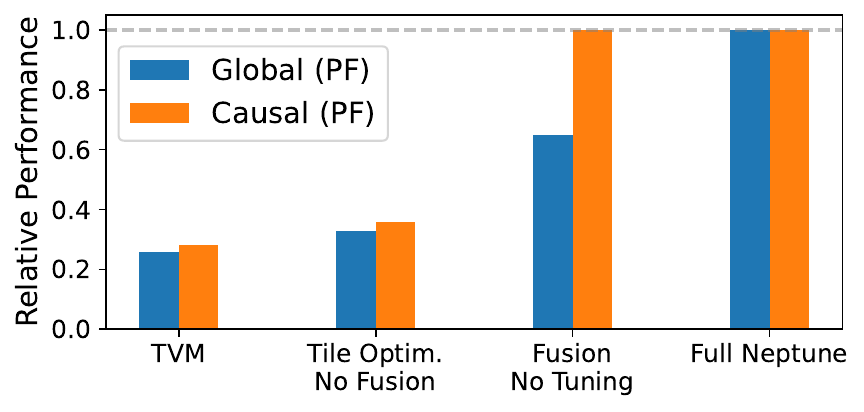}
    \caption{Ablation studies of the \NAME{} kernel performance
      on sequence length 512 on RTX 6000 Ada.
    The x-axis labels mark the component of \NAME{} we keep and remove.}
    \label{fig:ablation}
  \end{minipage}
\end{figure}

We also directly measure the number of registers, SMEM, and global memory
used in \NAME{}'s kernels.
On average, over all the Nvidia GPU setups,
\NAME{} uses $0.68 \times$ the registers of the baseline that uses the most registers,
and $47\%$ the SMEM of the baseline that uses the most SMEM.
The number of registers and the amount of SMEM mainly depend on the operator
and change little over input size.
We did not observe that they have \mbox{negative impact on speed.}

\subsection{Other Studies for Attention Models}

\textbf{Ablation Studies.} To understand the impact of each component of \NAME{},
we perform ablation studies on the performance of \NAME{}'s kernels.
We select the Global and Causal operators in prefill with sequence length 512 on RTX 6000 Ada.
Figure \ref{fig:ablation} shows the results of this experiment.
On the left, we start from baseline TVM with no \NAME{} components.
As we move to the right, we first add \NAME{}'s tile optimizations,
then \NAME{}'s fusion algorithms, and finally \NAME{}'s autotuner.
For both operators, fusion is the most impactful optimization,
and autotuning provides major improvement for Global and little for Causal.
Overall, the effect of the first three steps is consistent over operators and sequence lengths.

\noindent\textbf{Numerical Stability.}
We evaluate the numerical behavior of \NAME{}, Tri-Dao,
and FlexAttention kernels on the 10 attention operators.
\NAME{} produces kernels with competitive numerical precision.
Compared to a FP64 unfused attention kernel baseline,
\NAME{} kernels produce less or equal root mean square (RMS) error
than both Tri-Dao and FlexAttention kernels on all prefill operators.
\NAME{} also shows small RMS errors on decode operators.
The full \mbox{result is in Appendix \appref{sec:app:num-stab}.}

\subsection{Performance of \NAME{} on Non-Attention Operators}

Figure \ref{fig:extra-ops} presents the performance of \NAME{}
compared to PyTorch for the $L^2$ norm, RMSNorm, and performer operators.
The y-axis shows the relative performance of \NAME{} compared to PyTorch (higher
is better).
The x-axis shows the sequence length in the input tensor (log scale).

\begin{figure}[b]
  \begin{minipage}[t]{0.66\columnwidth}
    \centering
    \includegraphics[width=\columnwidth]{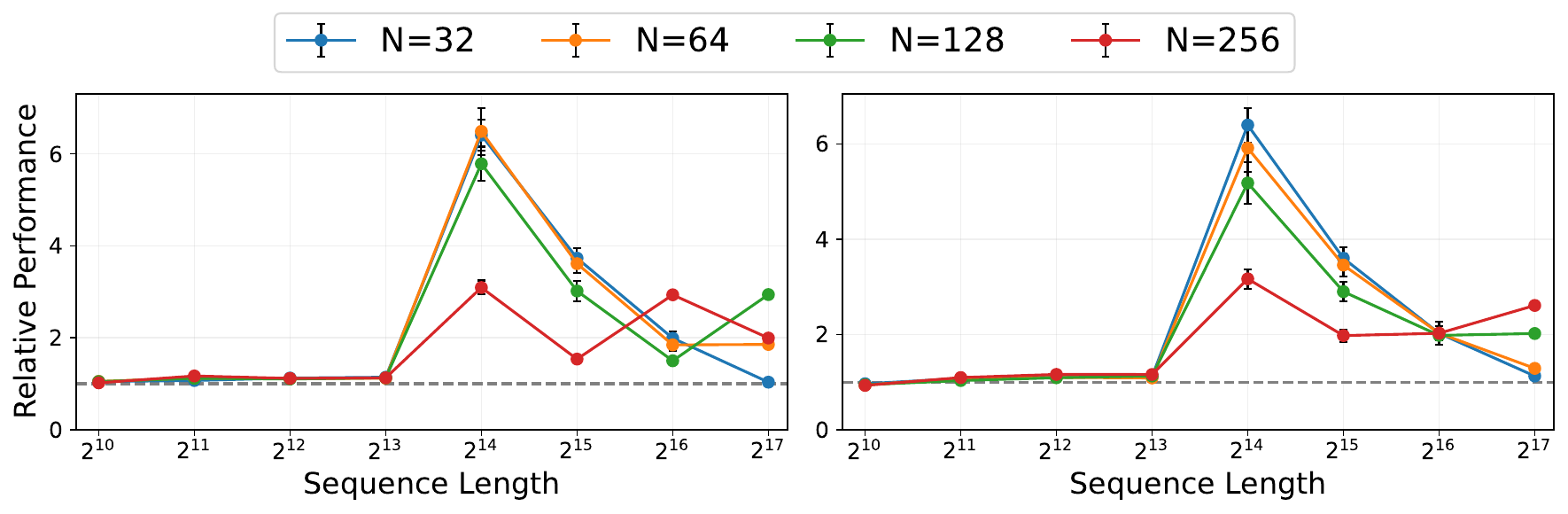}
  \end{minipage}
  \begin{minipage}[t]{0.325\columnwidth}
    \centering
    \includegraphics[width=\columnwidth]{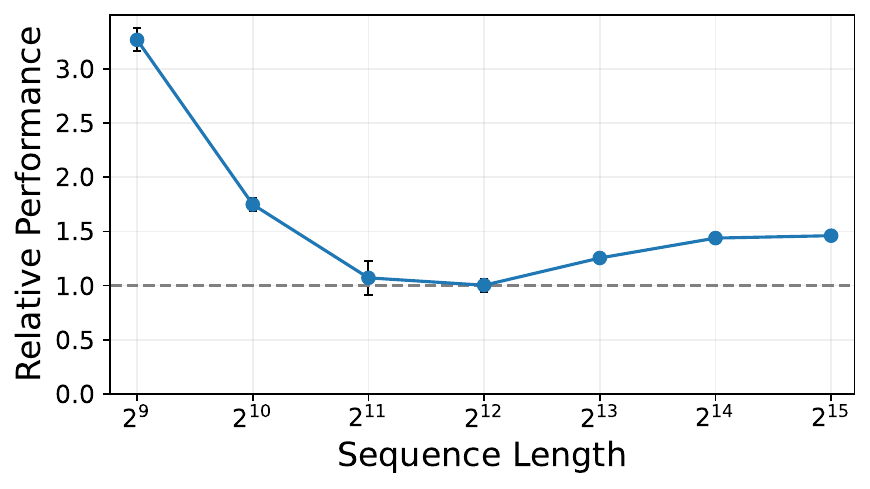}
  \end{minipage}
  \caption{Relative performance of \NAME{} kernels for $L^2$ norm (left),
  RMSNorm (middle), and performer (right), compared to PyTorch baseline. Higher is better.}
  \label{fig:extra-ops}
  \vspace{-.03in}
\end{figure}

For $L^2$ norm and RMSNorm, the sequence length is the number of columns $M$,
and multiple lines represent different numbers of rows $N$.
These two operators show similar performance trends with a mode switch.
On $M \leq 8192$, \NAME{} performs slightly better than PyTorch.
Here both \NAME{} and PyTorch use a single kernel to run the computation.
PyTorch produces a single kernel without deep fusion techniques like rolling update,
because the input is small enough for a threadblock to communicate entire rows
internally without going to global memory.
When $M > 8192$, \NAME{} starts to significantly outperform PyTorch,
as PyTorch now needs to use multiple kernels for the computation.
By fusing reductions together into a single kernel, \NAME{} avoids
two sources of overhead: kernel launch overhead and data movement overhead.
\NAME{}'s performance gain is the largest at $M = 16384$,
and decreases as kernel launch overhead becomes less significant.

For Performer, \NAME{} significantly outperforms PyTorch on smaller sequences ($L \leq 1024$).
The performance gain decreases as sequence length increases until $L = 2048$,
and increases again afterwards.
For all sequence lengths, \NAME{} fuses the entire performer operator into two kernels,
while PyTorch uses 17 kernels for the computation.
Compared to $L^2$ norm and RMSNorm, fusion is always beneficial for performer
due to the large number of computation steps. Similar to $L^2$ norm and RMSNorm,
the performance gain is larger for smaller sequences where kernel launch overhead dominates,
and data movement overhead becomes more significant as sequence length increases.
}

\section{Related Work}
\label{sec:rw}
{\textbf{Manual Tensor Operators.}
FlashAttention \cite{flash-attention} details an attention-specific fusion algorithm
with insights similar to \NAME{}'s general rolling update, and FlashDecoding \cite{flash-decoding}
proposes a fuse-and-split algorithm for decoding attention, similar to \NAME{}'s split-k update.
These works present transformed operators instead of a general means of transforming operators:
a library \cite{flash-attn-repo} offers kernels for common attention variants,
but specialized variants require manual implementation.
Manually implemented operators lack the flexibility to apply to diverse computations,
and may struggle to compose with automated compiler optimizations.

\textbf{Tile-Based Compilers.}
Many tensor compilers and programming frameworks decompose tensor-level operations into
operations over tensor tiles,
which often map directly to the underlying hardware.
Tile programming frameworks, such as Triton~\cite{triton},
Tilelang~\cite{tilelang}, and more~\cite{thunderkittens, pallas, cutlass},
provide tiles as a core abstraction and map them to target devices like GPU TensorCores.
Roller \cite{roller} recursively tiles kernels to match unknown accelerator architectures.
Tile compilers automate much of low-level intra-tile optimization,
but leave the burden of high-level optimization
like block tiling and operator fusion to the programmer.
As stand-alone systems, tile compilers function similarly to manual operators
at a slightly higher level of \mbox{abstraction, with similar limitations.}

\textbf{Scheduling Kernel Compilers.}
Halide~\cite{halide}, TVM~\cite{tvm}, and other compilers~\cite{tensorir, flextensor}
separate the specification of a tensor operator from its implementation, called schedule.
They search for schedules using autotuning or programmer input,
but the number of steps required to reach a high-performance schedule can overwhelm programmers.
Schedule primitives typically provide basic composable transformations,
but not algebra-rewriting \mbox{transformations like  \NAME{} provides.}

\textbf{Auto-Schedulers and Autotuners.}
Tensor compilers use autotuners to search for high-per\-for\-mance tensor programs on
specific hardware,
such as \cite{halide-2016, halide-2018, halide-2019} in Halide,
and AutoTVM and Ansor~\cite{ansor} in TVM.
Felix \cite{felix} provides a novel formulation of tensor program autotuning
as a continuous (gradient descent) search problem.
Triton \cite{triton} provides an enumerative search
while leaving the tuning knob definitions to the user.
Autotuning is also used to explore performance-accuracy trade-offs of deep learning
models~\cite{approxcaliper, approxtuner}.
Scheduling compilers also integrate with auto-schedulers to define search spaces for autotuners,
such as Ansor \cite{ansor} and MetaSchedule \cite{metaschedule} in TVM.
These auto-schedulers generate a list of candidate schedules for an autotuner to search from.
As \NAME{} provides operator fusion as new schedule primitives,
it can compose with auto-schedulers to \mbox{further automate schedule creation.}

\textbf{PyTorch-Based Optimizations.}
FlexAttention \cite{flexattn} is a PyTorch framework that
extends FlashAttention to allow customizing a mask and an element-wise score function.
It improves coverage for attention variants but is still limited to attention,
as it generates code from manual program templates.
In parallel to our work, Flashlight~\cite{flashlight} proposes semantic fusion to convert
two adjacent tensor kernels to a single kernel with online reduction.
This semantic fusion does not extend to more complex data flow like \NAME{} does,
and the design of the Flashlight system is \mbox{highly specific to PyTorch.}

\textbf{Graph- and Model-level Optimizations.}
Many tensor compilers like PyTorch \cite{pytorch}, XLA/HLO \cite{xla},
TVM \cite{tvm}, TASO \cite{taso}, and others \cite{halide, mirage, dnnfusion, tensorflow}
describe entire deep learning models as operator graphs and apply graph optimizations.
Operator graphs hide the detailed description per operator,
allowing optimizations to happen at a high level.
They are distinct from kernel-level optimizations in kernel compilers,
although they can co-exist in the same framework as a form of hierarchical optimization.
Graph-level frameworks can provide operator fusion \cite{evt, dnnfusion}
and algebraic rewriting \cite{taso, taso2, mlsys_egraph_tensor, halide,
halide_rewrite_synthesis, pet}.
However, they do not support advanced operator fusion in \NAME{} or allow them to be
expressed as rewrite rules.
Operator graphs are not the ideal abstraction for \NAME{}'s operator fusion,
which requires loop fusion and tile-level rewriting to occur in tandem.

\textbf{Superoptimization Compilers.}
Some tensor compilers \cite{taso, superopt-eqsat, superopt-pet, superopt-unity} apply
superoptimization,
which transforms the given program in potentially semantic-breaking ways,
and impose additional checks to select correct implementations.
Most superoptimizers transform a single level of abstraction at a time
(i.e. only graph rewrites or only instruction rewrites),
unlike advanced fusions in \NAME{} which are expressed
as a combination of loop and algebra transformations.
As an exception, Mirage \cite{mirage} is a hierarchical superoptimizer that
jointly transforms the program at thread, tile, and operator levels.
Mirage uses a probabilistic correctness test on integer inputs,
producing programs that have (high) probability of being correct,
contrasting with \NAME{}'s formal correctness.
The test does not account for the numerical error on floating-point inputs,
and Mirage has visible numerical instability in some cases such as attention.
The search algorithm in Mirage can take minutes to hours,
and may end up with no valid kernels
or find kernels with unpredictable performance (as observed in experiments in \S\ref{sec:eval}).

\textbf{Nondeterministic/Approximate Dependency Breaking.}
Some works apply dependency-breaking transformations, e.g.,  Hogwild~\cite{hogwild}
for distributed neural network training and Deiana et al.~\cite{campanoni} for
thread-level parallelization.
These works leverage application-level tolerance to error or nondeterminism to break
performance-limiting dependencies.
Hogwild declines to repair broken dependencies, creating an approximate result, while
Deiana et al.~use runtime repair
suitable for CPU workloads but infeasible for deterministic tensor workloads.
\NAME{} differs from these works as it provides
static transformations that recover equivalence {(on real numbers) at compile time.}
}

\section{Conclusion}
\label{sec:conclusion}
{We presented \NAME{}, a novel tensor compiler for advanced ML operator fusion.
\NAME{} shows, for the first time, that advanced fusion algorithms can be included in a
standard compiler schedule,
to transform plain attention into kernels like FlashAttention and FlashDecoding,
reaching and exceeding state-of-the-art kernel performance.
Our subsequent work Nautilus~\cite{nautilus} shows the benefits of our transformations on
new GPU architectures (Hopper and Blackwell)
and automates scheduling.
\NAME{} opens a new direction towards bringing such optimizations
fully within the scope of optimizing tensor compilers.
We anticipate that this integration will pave the way for automatic discovery of new
efficient advanced ML operators on a broad range \mbox{of emerging AI hardware platforms.}

\vspace{5pt}

\subsection*{Acknowledgments}
We thank our anonymous reviewers for the feedback.
This research was supported in part by the NSF grant No. CCF-2217144 and the IBM-Illinois
Discovery Accelerator Institute.
This research used DeltaAI advanced computing and data resource,  supported by the NSF
(award OAC 2320345) and the State of Illinois.
}

{
  \bibliographystyle{ACM-Reference-Format}
  \bibliography{references}
}

\newpage
\appendix
{\clearpage
\section*{Appendix}
\vspace{1in}
\section{\NAME{} Inputs}
\label{sec:app:neptune-inputs}

Neptune takes as input a compute definition of the operator to optimize,
and a schedule that describes how to optimize it.

Figure \ref{fig:appendix-attention} shows one possible \NAME{} input, a flexible compute
definition for attention.
It allows customizing bias and mask conditions,
and in 38 lines of code it covers all operators in Table \ref{tab:operators}
except GQA operators.

Figure \ref{fig:appendix-schedule} shows the schedule that pairs with the compute definition.
In 28 lines of code, it applies rolling update to fuse the attention computation
into a single loop nest, and produces many of the high-performance kernels in our evaluation.

\begin{figure}[ht]
  \centering
  \begin{minted}{python}
def create_general_attention(
    B: int, N: int, QS: int, KVS: int, H: int, mask_cond: Callable, score_mod: Callable
):
  q = placeholder((B, N, QS, H), "float16", name="q")
  k = placeholder((B, N, KVS, H), "float16", name="k")
  v = placeholder((B, N, KVS, H), "float16", name="v")
  p = batch_matmul(q, k, trans_b=True, out_dtype="float32")
  score = compute(
      p.shape, lambda *ax: if_then_else(mask_cond(*ax), score_mod(p(*ax), *ax), float("-inf")),
      name="score_mod")
  j = reduce_axis((0, KVS), name="j")
  s_max = compute(
      (B, N, QS), lambda b, n, i: max(score(b, n, i, j), axis=j), name="softmax_maxelem")
  s_exp = compute(
      (B, N, QS, KVS), lambda b, n, i, j: exp(score(b, n, i, j) - s_max(b, n, i)),
      name="softmax_exp")
  s_expsum = compute(
      (B, N, QS), lambda b, n, i: sum(s_exp(b, n, i, j), axis=j), name="softmax_expsum")
  s_exp = compute(
      p.shape, lambda *axes: s_exp(*axes).astype(q.dtype), name="softmax_exp_f16")
  sv = batch_matmul(s_exp, v, trans_b=False, out_dtype="float32")
  sv = compute(
      sv.shape, lambda b, n, i, j: sv(b, n, i, j) / s_expsum(b, n, i), name="softmax_norm")
  return compute(sv.shape, lambda *i: sv(*i).astype(q.dtype), "cast")
  \end{minted}
  \caption{Compute definition for the attention operator, which \NAME{} takes as input.
  This definition covers all operators in Table \ref{tab:operators} except GQA operators.}
  \label{fig:appendix-attention}
\end{figure}

\begin{figure}[ht]
  \centering
  \begin{minted}{python}
def schedule_attn_with_rolling_update(sch: Schedule):
  b0 = sch.get_block("batch_matmul_1")
  b1 = sch.get_block("T_score_mod")
  b2 = sch.get_block("T_softmax_maxelem")
  b3 = sch.get_block("T_softmax_exp")
  b4 = sch.get_block("T_softmax_exp_cast")
  b5 = sch.get_block("T_batch_matmul_NN")
  b6 = sch.get_block("T_softmax_expsum")
  b7 = sch.get_block("T_softmax_norm")
  b8 = sch.get_block("T_cast")
  *axes, i, j, k = sch.get_loops(b0)
  i0, j0 = sch.tile([i, j], [128, 32])
  sch.compute_at(sch.cache_read(b0, 0, "shared"), i0)
  sch.bind_block_idx([*axes, i0], ["blockIdx.x", "blockIdx.y", "blockIdx.z"])
  sch.reverse_compute_at(b1, j0)
  b2rf = sch.rolling_update(b2, j0, factor_axis=0)
  b6rf = sch.rolling_update(b6, j0, factor_axis=0)
  b5rf = sch.rolling_update(b5, j0, factor_axis=0)
  sch.reverse_compute_at(b7, i0)
  sch.reverse_compute_at(b8, i0)
  for blk in [b0, b1, b2, b2rf, b3, b4, b5, b5rf, b6, b6rf, b7]:
      sch.set_scope(blk, 0, "shared")
  sch.split_scan_buffer(b2, j0, 0)
  sch.decompose_reduction(b5, j0)
  sch.decompose_reduction(b6, j0)
  \end{minted}
  \caption{Schedule for the attention operator, which \NAME{} takes as input.
  This schedule pairs with the compute definition in Figure \ref{fig:appendix-attention}.}
  \label{fig:appendix-schedule}
\end{figure}

\clearpage
\section{Step-By-Step Algorithm Application on Motivational Example}
\label{sec:app:step-by-step}

Here we include the detailed steps of our rolling update algorithm for the example from
Figure~\ref{fig:example-orig}.

\subsection{Rolling Update}
\def\inpij{\texttt{inp[i, j]}}
\begin{figure}[!h]
  \centering
  \begin{subfigure}[b]{0.3\textwidth}
    \centering
    \small
    \begin{minted}{python}
# s_max:
for i in range(2):
  for j in range(4):  # loop_j
    xmax[i] = max(
      xmax[i], inp[i, j])
# s_sum:
for i, j in grid(2, 4)
  xsum[i] = xsum[i] + exp(
    inp[i, j] - xmax[i])
    \end{minted}
    \caption{Step 1: dataflow reorganization.
    This step inlines \sexp{} (into \ssum{}) and finds $\redpred = \{\smax\}$.}
    \label{fig:example-step1}
  \end{subfigure}
  \hspace{1pt}
  \begin{subfigure}[b]{0.3\textwidth}
    \centering
    \small
    \begin{minted}{python}
for i in range(2):
  for j in range(4):  # loop_j
    # s_max:
    xmax[i] = max(
      xmax[i], inp[i, j])
    # s_sum:
    xsum[i] = xsum[i] + exp(
      inp[i, j] - xmax[i])
    \end{minted}
    \caption{Step 2: loop transformation. Naive loop fusion fuses \ssum{} and \smax{}
    under \loopj{}.}
    \label{fig:example-step2}
  \end{subfigure}
  \hspace{1pt}
  \begin{subfigure}[b]{0.37\textwidth}
    \includegraphics[width=\textwidth,trim=0mm 5mm 2mm 0mm,clip]{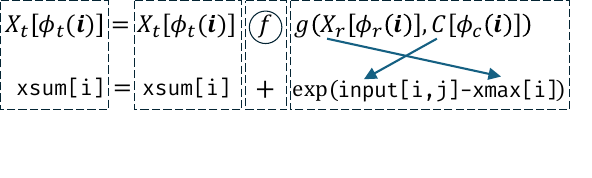}
    \small
    \begin{align*}
      X_t & = \xsum{}; & X_r     & = \xmax{}; \quad C = \inp{} \\
      f   & = +;       & g(x, c) & = \exp(c - x)
    \end{align*}
    \caption{Step 3: pattern matching on the compute statement of $L_t = \ssum{}$.}
    \label{fig:example-step3}
  \end{subfigure}
  \begin{subfigure}[b]{0.48\textwidth}
    \footnotesize
    \begin{align*}
      y = g(x, c) & = \exp(c - x)\ \Rightarrow\ c = g_c^{-1}(x, y) = x + \ln(y) \\
      h(t, r, r') & = g(r', r + \ln(t)) = \exp(r - r') \cdot t
    \end{align*}
    Proving commutativity with $f$:
    \[
      h(x + y, r, r') = \exp(r - r') \cdot (x + y) = h(x, r, r') + h(y, r, r')
    \]
    \caption{Step 4: finding the repair function $h$. Find the inverse of $g$,
      use Equation \ref{eq:h-solution} to find $h$,
    and prove Equation \ref{eq:h-conditions-2} (commutativity with $f$) for $h$.}
    \label{fig:example-step4}
  \end{subfigure}
  \hspace{1pt}
  \begin{subfigure}[b]{0.48\textwidth}
    \footnotesize
    \vspace{3pt}
    For \smax{}, cache its output \xmax{} for the previous iteration:
    \[
      \xmaxprevi := \tagged{\xmaxi}{j - 1}; \quad
      \xmaxcurri := \tagged{\xmaxi}{j}
    \]
    Apply $h(t, r, r')$ with $r = \xmaxprevi$ and $r' = \xmaxcurri$:
    \begin{align*}
      \fixed{\xsumi} & = \exp(\xmaxprevi - \xmaxcurri) \cdot \xsumi \\
      & + \exp(\inpij - \xmaxcurri)
    \end{align*}
    \caption{Step 5: applying the repair function.}
    \label{fig:example-step5}
  \end{subfigure}
  \caption{All intermediate results of rolling update, when applied to Figure
  \ref{fig:example-orig}.}
  \label{fig:example-assemble}
\end{figure}

\subsection{Privatization}
Figure~\ref{fig:example-priv} presents privatization combined with rolling update in \NAME{}.
On top of the rolling update output in Figure \ref{fig:example-ours},
we privatize \smax{} and \ssum{} with a split size of $2$.
\texttt{xmax\_1p} and \texttt{xsump} are the local arrays.

\begin{figure*}[h]
  \centering
  \small
  \begin{minted}{python}
for i in range(2):
  xmax_0[i] = -inf
  for j1 in range(2):
    for j2 in range(2):
      # max_local
      xmax_1p[i, j1] = max(xmax_1p[i, j1], inp[i, j1 * 2 + j2])
    # max_global
    xmax_1[i] = max(xmax_1[i], xmax_1p[i, j1])
    for j2 in range(2):
      # sum_local
      xsump[i, j1] += exp(inp[i, j1 * 2 + j2] - xmax_1[i])
    xsum[i] = exp(xmax_0[i] - xmax_1[i]) * xsum[i] + xsump[i, j1]
    xmax_0[i] = xmax_1[i]
    \end{minted}
  \vspace{-.1in}
  \caption{
    Privatization combined with rolling update in \NAME{}.
  }
  \label{fig:example-priv}
\end{figure*}

\clearpage

\section{Additional Example Operators Amenable to Neptune Transformations}
\label{sec:app:example-operators}

\hspace{\parindent}\textbf{Attention definition.}
The Attention operator is a sequence of four compute steps: matrix multiplication (matmul),
element-wise score computation (division), softmax, and another matmul.
Attention typically use a numerically stable softmax implementation,
subtracting the input by per-row maximum to prevent exponential overflow.
Equation \ref{eq:attention} shows the compute steps of Attention over 2D inputs
$q, k, v \in \mathbb{R}^{L \times D}$.
$i$ and $j$ are indices along the $L$ (sequence length) dimension,
and $d$ is the index along the $D$ (number of features) dimension.
\begin{align}
  p_{ij} &= \sum_d q_{id} v_{jd}; \quad p'_{ij} = \frac{p_{ij}}{\sqrt{D}} &
  &\text{matmul, division} \nonumber \\
  m_i &= \max_j p'_{ij}; \quad e_{ij} = \exp(p'_{ij} - m_i); \quad s_i = \sum_j e_{ij} &
  &\text{softmax (steps 1-3)} \label{eq:attention} \\
  o_{id} &= \sum_j e_{ij} v_{jd}; \quad o'_{id} = \frac{o_{id}}{s_i} &
  &\text{matmul, softmax (step 4)} \nonumber
\end{align}

\textbf{Performers.}
Performers~\cite{performers} is an alternative transformer architecture that uses
non-softmax computation to approximate softmax.
The performer operator takes three inputs $Q$, $K$, and $V$
with a sequence length ($L$) and a number of features ($D$) dimension,
similar to attention.
Performer applies a randomly sampled projection matrix of shape $M \times D$
along with certain normalization on $Q$ and $K$
to produce $Q', K' \in \mathbb{R}^{L \times M}$.
Then performer computes $A = Q' K' V$ and applies additional normalization on $A$
without needing to compute a softmax.
These normalization steps provide multiple reductions that \NAME{}'s reduction fusion
applies to.

\NAME{} fuses performer into two loop nests, applying reduction fusion four times.
Two fusions have computation pattern $g_1$,
and the other two have distinct computation patterns $g_2$ and $g_3$:
\begin{align*}
  g_1(r_1, c_1, c_2) = \alpha \exp(c_1 - r_1 - \beta c_2);&&
  &h_1(t, r, r') = t \exp(r' - r) \\
  g_2(r_1, c_1, c_2, c_3) = \alpha c_3 \exp(c_1 - r_1 - \beta c_2);&&
  &h_2(t, r, r') = t \exp(r' - r) \\
  g_3(r_1, r_2, c_1, c_2, c_3) = \alpha \exp(c_1 - r_1 - \beta c_2 c_3) / r_2;&&
  &h_3(t, r_1, r_1', r_2, r_2') = t (r_2 / r_2') \exp(r_1 - r_1')
\end{align*}
$\alpha$ and $\beta$ are compile-time constants.
The reduction function in all these cases is $f(x, y) = x + y$.
$g_1$, $g_2$, and $g_3$ are all different from any softmax-based attention variants.
The solution that \NAME{} finds is also given as $h_1$, $h_2$, and $h_3$.
The last case $g_3$ is a fusion between one reduction and two producers,
as the two reduction inputs $r_1$ and $r_2$ indicate,
which \NAME{} is also capable of (\S\ref{sec:rolling-update},
Eqns.~\ref{eq:h-cond-1-general} and \ref{eq:h-cond-2-general}).

Upstream optimizations can change the program that reaches \NAME{} fusion.
One optimization is to delay the normalization on $A = Q' K' V$
to the end of the computation.
This optimization can improve the performance and numerical stability of \NAME{}-fused kernel,
as the normalization involves a division that is now delayed outside of the rolling loop
and only applied once.
When this optimization is applied, the fourth fusion ($g_3$)
will have an identical compute pattern as the third one ($g_2$).

\clearpage
\section{Proofs of Rolling Update Correctness Theorems}
\label{sec:app:proofs}

\def\jf{\incorrect{\bj_f}}
\def\jt{\fixed{\bj_t}}
\newcommand{\g}[2]{g\left(\tagged{X_r}{#1}, C_{#2}\right)}
\newcommand{\rterm}[2]{\tagged{R}{#1}_{#2}}

We use these abbreviations for the following proofs
\[
  \access{C}{c}{\bi, \bj'} =: C_{\bj'} \quad
  \tagged{\access{X_t}{t}{\bi}}{\bj} =: \tagged{X_t}{\bj} \quad
  \tagged{\access{X_r}{r}{\bi}}{\bj} =: \tagged{X_r}{\bj}.
\]
essentially dropping the tensor access function and the map iteration vector $\bi$,
because all the proofs involve a single loop nest where $\bi$ is fixed.

\begin{proof}[Proof of Lemma \ref{lemma:h-tag-update}]
  \def\xrfrom{\tagged{X_r}{\jf}}
  \def\xrto{\tagged{X_r}{\jt}}

  We first show that the constructive solution of $h$, provided in Eq.~\ref{eq:h-solution},
  is a solution to condition Eq.~\ref{eq:h-conditions-1}.
  Since $g_c^{-1}$ is the second-argument inverse of $g$, we have
  \[
    g_c^{-1}(x, g(x, c)) = c
  \]
  In Eq.~\ref{eq:h-solution}, substitute $t = g(r, c)$ to get
  \[
    h(g(r, c), r, r') = g(r', g_c^{-1}(r, g(r, c))) = g(r', c)
  \]
  which is exactly Eq.~\ref{eq:h-conditions-1}.
  Next, given that $h$ satisfies the conditions Eqns.~\ref{eq:h-conditions-1} and
  \ref{eq:h-conditions-2},
  we prove that $h$ satisfies Def.~\ref{def:h-tag-update} which is an equality.
  The left-hand side of the equality is $h$ applied to the reduce expression
  (rewritten using our abbreviations)
  \[
    \fold\left(f, 0 \preceq \bj' \preceq \bj_0, \g{\jf}{\bj'}\right) =: \rterm{\jf}{\bj_0}
  \]
  and the right-hand side is the updated reduce expression
  \[
    \fold\left(f, 0 \preceq \bj' \preceq \bj_0, \g{\jt}{\bj'}\right) =: \rterm{\jt}{\bj_0}
  \]
  The subscript $\bj_0$ in $R_{\bj_0}$ indicates the upper bound of the reduce range.
  Therefore, we are to prove
  \[
    \forall \bj_0, \jf, \jt \in \dom{s}, \jf \preceq \jt \preceq \bj_0 \quad
    h\left(\rterm{\jf}{\bj_0}, \xrfrom, \xrto\right) = \rterm{\jt}{\bj_0}
  \]
  which we prove by induction over $\bj_0$.
  We introduce $\bj$ as the induction variable to go from $0$ to $\bj_0$,
  while all tags are fixed, and maintain the inductive hypothesis
  \begin{equation}
    \forall 0 \preceq \bj \preceq \bj_0 \quad
    h\left(\rterm{\jf}{\bj}, \xrfrom, \xrto\right) = \rterm{\jt}{\bj}
    \label{eq:h-tag-update-inductive}
  \end{equation}

  \newcommand{\rfrom}[1]{\rterm{\jf}{#1}}
  \newcommand{\rto}[1]{\rterm{\jt}{#1}}
  \newcommand{\gfrom}[1]{\g{\jf}{#1}}
  \newcommand{\gto}[1]{\g{\jt}{#1}}
  \newcommand{\fh}[1]{h\left(#1, \xrfrom, \xrto\right)}
  \begin{itemize}
    \item Base case: when $\bj = 0$, $\rfrom{0}$ and $\rto{0}$ are each a single term,
      so we spell out these terms in the inductive hypothesis to get
      \[
        h\left(\gfrom{0}, \xrfrom, \xrto\right) = \gto{0}
      \]
      which is true by applying Eq.~\ref{eq:h-conditions-1}.
    \item Inductive step: for any $\bj$ where the inductive hypothesis
      Eq.~\ref{eq:h-tag-update-inductive} is true,
      we move on to $\nextj$ by adding one term to $\rfrom{\bj}$:
      \[
        \rfrom{\nextj} = \rfrom{\bj} \cf \gfrom{\nextj}
      \]
      Apply $h$ to both sides, apply Eq.~\ref{eq:h-conditions-2},
      then apply Eq.~\ref{eq:h-conditions-1},
      and finally apply the inductive hypothesis on the underlined term, to get
      \begin{align*}
        & \fh{\rfrom{\nextj}} = \fh{\rfrom{\bj} \cf \gfrom{\nextj}} \\
        =\  & \fh{\rfrom{\bj}} \cf \fh{\gfrom{\nextj}}                  \\
        =\  & \underline{\fh{\rfrom{\bj}}} \cf \gto{\nextj}             \\
        =\  & \rto{\bj} \cf \gto{\nextj} = \rto{\nextj}
      \end{align*}
      Therefore, the inductive hypothesis is true for $\nextj$.
  \end{itemize}
  By induction, the theorem is true for all $\bj \preceq \bj_0$, and therefore true for $\bj_0$.
\end{proof}

\newcommand{\xt}[1]{\tagged{X_t}{#1}}
\def\xrp{\tagged{X_r}{\prevj}}
\def\xr{\tagged{X_r}{\bj}}
\def\xrn{\tagged{X_r}{\nextj}}

\begin{proof}[Proof of Theorem \ref{thm:rollup-correct}]
  \newcommand{\ggeq}[1]{\g{#1}{#1}}
  The goal of this proof is to show that the result of rolling update transformed program
  matches that of the original program.
  We copy the explicit expression of $X_t$ in the original program
  from Eq.~\ref{eq:original-explicit}, and apply our abbreviations:
  \begin{equation}
    \original{\xt{\bm}} = \fold\left(f, 0 \preceq \bj' \preceq \bm, \g{\bm}{\bj'}\right)
    \label{app:eq:original-explicit}
  \end{equation}

  On the other hand, the repaired program Eq.~\ref{eq:fixed-recurrent} is (with tag added
  for clarity)
  \[
    \xt{\bj} = h\left(\xt{\prevj}, \xrp, \xr\right) \cf \ggeq{\bj}
  \]
  and we assert it produces this value at iteration $\bj$:
  \begin{equation}
    \fixed{\xt{\bj}} = \fold\left(f, 0 \preceq \bj' \preceq \bj, \g{\bj}{\bj'}\right) =:
    \rterm{\bj}{\bj}
    \label{app:eq:fixed-explicit}
  \end{equation}
  If this is true,
  then we have $\fixed{\xt{\bm}} = \original{\xt{\bm}}$
  by substituting $\bm$ for $\bj$ in Eq.~\ref{app:eq:fixed-explicit}
  and comparing to Eq.~\ref{app:eq:original-explicit}.
  We prove this Eq.~\ref{app:eq:fixed-explicit} by induction on $\bj$.
  \begin{itemize}
    \item Base case: when $\bj = 0$, the recurrent $X_0$ has iterated once:
      \[
        \xt{0} = \ggeq{0}
      \]
      and the explicit $\fixed{\xt{0}}$ has a single term:
      \[
        \rterm{0}{0} = \ggeq{0}
      \]
      and they are equal, so the inductive hypothesis is true.
    \item Inductive step: for any $\bj$ where the inductive hypothesis is true,
      the next iteration $\nextj$ updates the recurrent $X_t$ once.
      We write down the updated value,
      apply the inductive hypothesis to convert to explicit form,
      then use the tag-updating property of $h$ to get
      \begin{align*}
        & \xt{\nextj}                                                   \\
        & = h\left(\xt{\bj}, \xr, \xrn\right) \cf \ggeq{\nextj}         \\
        & = h\left(\rterm{\bj}{\bj}, \xr, \xrn\right) \cf \ggeq{\nextj} \\
        & = \rterm{\nextj}{\bj} \cf \ggeq{\nextj}                       \\
        & = \rterm{\nextj}{\nextj}
      \end{align*}
      Therefore, the inductive hypothesis is true for $\nextj$.
  \end{itemize}
  By induction, Eq.~\ref{app:eq:fixed-explicit} is true for all $\bj \preceq \bm$.
  Substituting $\bm$ for $\bj$, we get
  \[
    \fixed{\xt{\bm}} = \fold\left(f, 0 \preceq \bj' \preceq \bm, \g{\bm}{\bj'}\right)
  \]
  which matches Eq.~\ref{app:eq:original-explicit}, and therefore proves the theorem.
\end{proof}

\newcommand{\loc}[1]{#1^\text{local}}
\newcommand{\glb}[1]{#1^\text{global}}
\def\xtl{X_{t,l}}
\def\xrl{X_{r,l}}
\def\xtg{X_{t,g}}
\def\xrg{X_{r,g}}

\begin{proof}[Proof of Theorem \ref{thm:su-correct}]
  The goal of this proof is to show that the result of split-k update transformed program
  matches that of the original program.
  The expected expression of $X_t$ in the original program is the same as in
  Eq.~\ref{app:eq:original-explicit}:
  \[
    \original{\xt{\bm}} = \fold\left(f, 0 \preceq \bj \preceq \bm, \g{\bm}{\bj}\right)
  \]
  Privatization splits this reduction into a local reduction,
  that produces $\bm_0$ partial results, each result reduced from $\bm_1$ terms.
  Iterators for $\bm_0$ and $\bm_1$ combine to an iterator for $\bm$
  similar to how loop tiling works:
  \[
    \forall 0 \preceq \bj_0 \preceq \bm_0, 0 \preceq \bj_1 \preceq \bm_1 \quad
    \exists 0 \preceq \bj \preceq \bm \quad
    \bj_0 \cdot \bm_1 + \bj_1 = \bj
  \]

  The result of the local reduction is defined by the local reduction that produces it:
  \[
    \xtl[\bj_0] = \xtl[\bj_0] \cf g\left(\xrl[\bj_0], C_{\bj_0 \cdot \bm_1 + \bj_1}\right)
  \]
  $\bj_0$ shows up in the index of tensor $\xtl$ because the local reduction produces
  $\bj_0$ partial results.
  We convert this program into the following expression for $\xtl$:
  \[
    \xtl[\bj_0] = \fold\left(f, 0 \preceq \bj_1 \preceq \bm_1, g\left(\xrl[\bj_0],
    C_{\bj_0 \cdot \bm_1 + \bj_1}\right)\right)
  \]
  The global reduction further combines these partial results into a single result
  over $\bj_0$ iterations:
  \[
    \xtg = \xtg \cf h(\xtl[\bj_0], \xrl[\bj_0], \xrg)
  \]
  We substitute the expression of $\xtl[\bj_0]$ into this program to get
  \[
    \xtg = \xtg \cf h(\fold\left(f, 0 \preceq \bj_1 \preceq \bm_1, g\left(\xrl[\bj_0],
    C_{\bj_0 \cdot \bm_1 + \bj_1}\right)\right), \xrl[\bj_0], \xrg)
  \]
  Using the tag-updating property of $h$, the expression that we expanded from $\xtl[\bj_0]$
  drops all $\xrl[\bj_0]$ and replaces them with $\xrg$:
  \[
    \xtg = \xtg \cf \fold\left(f, 0 \preceq \bj_1 \preceq \bm_1, g\left(\xrg, C_{\bj_0
    \cdot \bm_1 + \bj_1}\right)\right)
  \]
  which has the following explicit expression:
  \begin{align*}
    \xtg & = \fold\left(f, 0 \preceq \bj_0 \preceq \bm_0, \fold\left(f, 0 \preceq \bj_1
    \preceq \bm_1, g\left(\xrg, C_{\bj_0 \cdot \bm_1 + \bj_1}\right)\right)\right) \\
    & = \fold\left(f, 0 \preceq \bj \preceq \bm, g\left(\xrg, C_{\bj}\right)\right)
  \end{align*}
  which matches the expected expression of $\xtg$ in the original program.
\end{proof}

\clearpage
\section{Additional Evaluation}
\label{sec:app:add-eval}

\subsection{Compilation Statistics}
We provide some statistics \NAME{}'s compilation pipeline
when generating kernels we have evaluated in Section \ref{sec:eval}.
For these evaluated operators,
\NAME{} with 128 autotuning iterations takes 1.5 to 10 minutes to run on each setup.
Triton-based tile optimization takes 50\% to 90\% of the time (depending on the input shape).
\NAME{}'s template-guided optimization (rolling/split-k update) takes
5\% to 10\% of the time (does not depend on input shape),
and the rest is TVM autotuning search algorithm.

\subsection{Operator Evaluation}

\begin{figure*}[h]
  \centering
  \begin{subfigure}[b]{0.95\textwidth}
    \includegraphics[width=\textwidth]{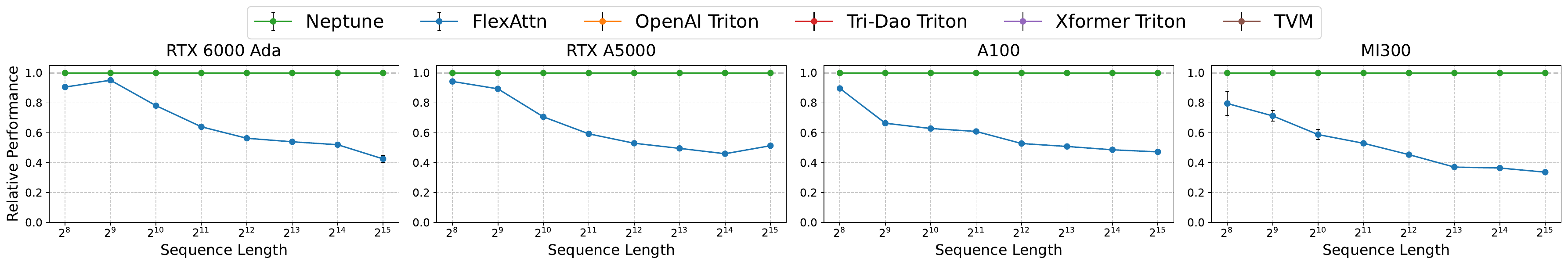}
    \vspace{-1.7em}
    \caption{ALiBi (PF) operator. Optimized with rolling update.}
  \end{subfigure}
  \begin{subfigure}[b]{0.95\textwidth}
    \includegraphics[width=\textwidth]{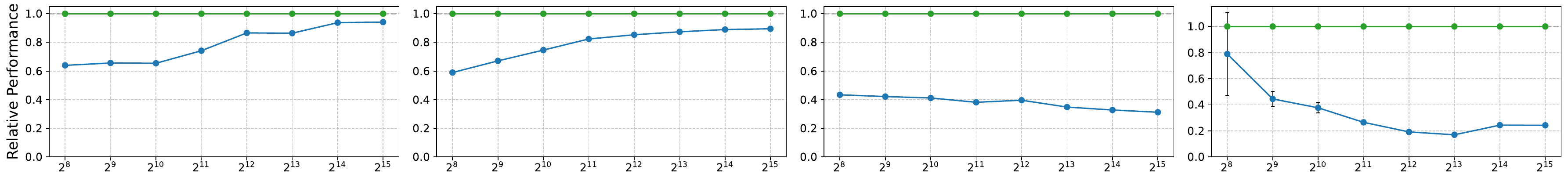}
    \vspace{-1.7em}
    \caption{ALiBi (DC) operator. Optimized with split-k update.}
  \end{subfigure}
  \begin{subfigure}[b]{0.95\textwidth}
    \includegraphics[width=\textwidth]{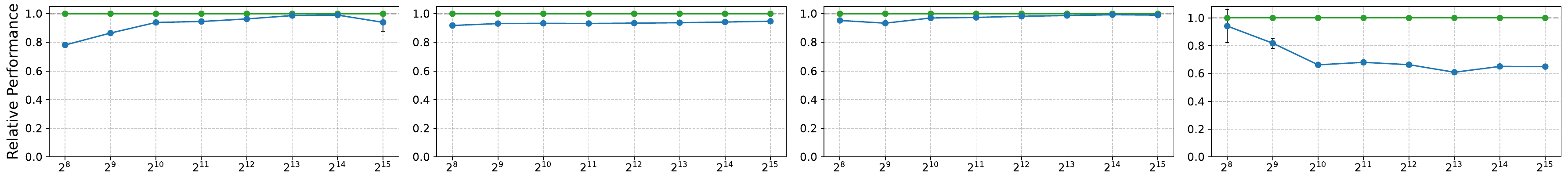}
    \vspace{-1.7em}
    \caption{SoftCap (PF) operator. Optimized with rolling update.}
  \end{subfigure}
  \begin{subfigure}[b]{0.95\textwidth}
    \includegraphics[width=\textwidth]{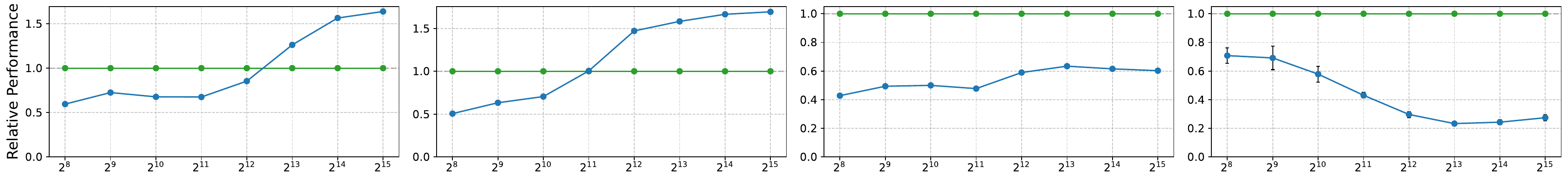}
    \vspace{-1.7em}
    \caption{SoftCap (DC) operator. Optimized with split-k update.}
  \end{subfigure}
  \begin{subfigure}[b]{0.95\textwidth}
    \includegraphics[width=\textwidth]{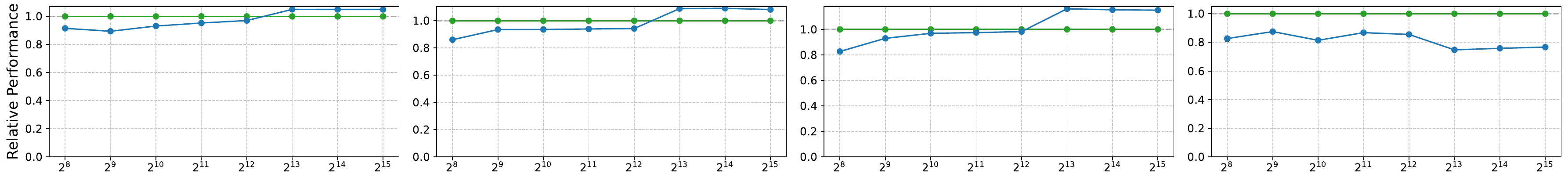}
    \vspace{-1.7em}
    \caption{Window (PF) operator. Optimized with rolling update.}
  \end{subfigure}
  \caption{Performance of \NAME{} vs.~other tensor compilers
  on the five operators not shown in Figure \ref{fig:rq1-first5-ops}.}
\end{figure*}

\clearpage

\begin{figure*}[t]
  \centering
  \begin{subfigure}[b]{0.95\textwidth}
    \includegraphics[width=\textwidth]{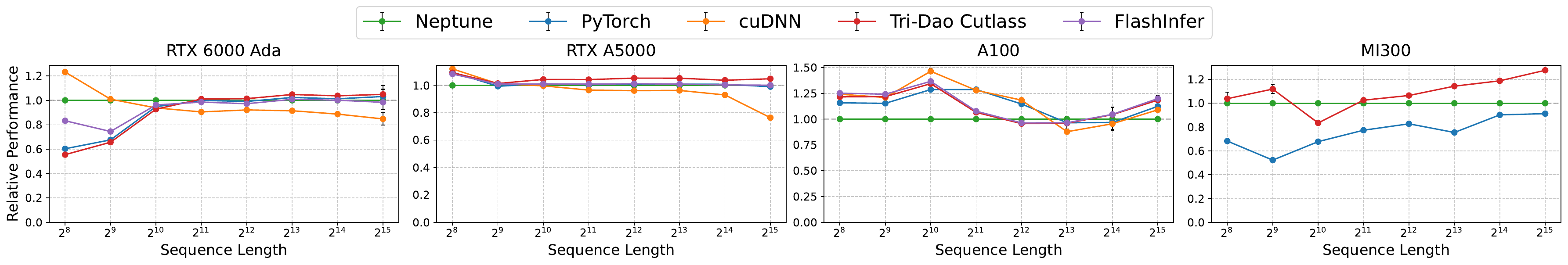}
    \vspace{-1.7em}
    \caption{Global (PF) operator. Optimized with rolling update.}
  \end{subfigure}
  \begin{subfigure}[b]{0.95\textwidth}
    \includegraphics[width=\textwidth]{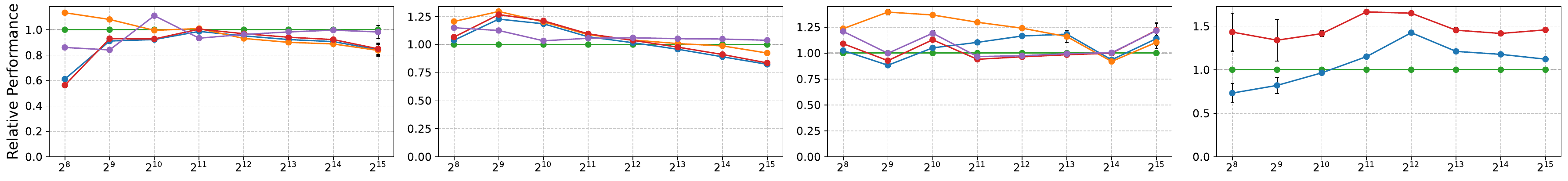}
    \vspace{-1.7em}
    \caption{Causal (PF) operator. Optimized with rolling update.}
  \end{subfigure}
  \begin{subfigure}[b]{0.95\textwidth}
    \includegraphics[width=\textwidth]{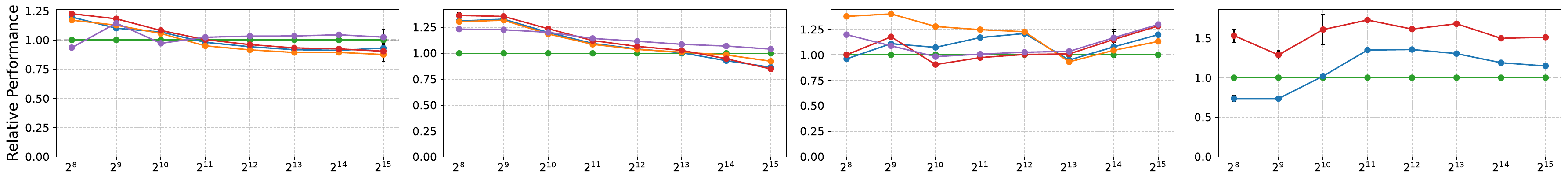}
    \vspace{-1.7em}
    \caption{GQA (PF) operator. Optimized with rolling update.}
  \end{subfigure}
  \begin{subfigure}[b]{0.95\textwidth}
    \includegraphics[width=\textwidth]{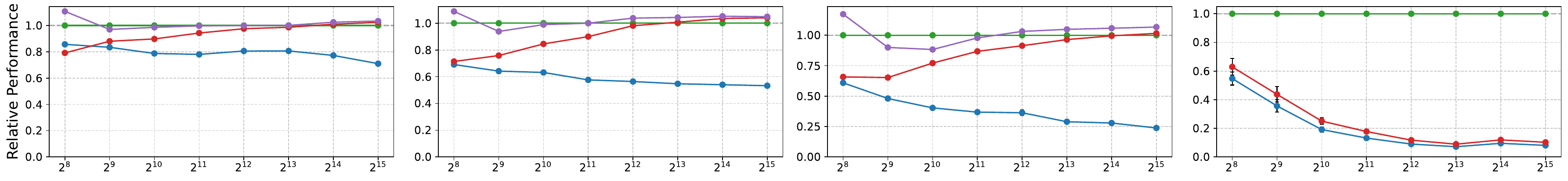}
    \vspace{-1.7em}
    \caption{Causal (DC) operator. Optimized with split-k update.}
  \end{subfigure}
  \begin{subfigure}[b]{0.95\textwidth}
    \includegraphics[width=\textwidth]{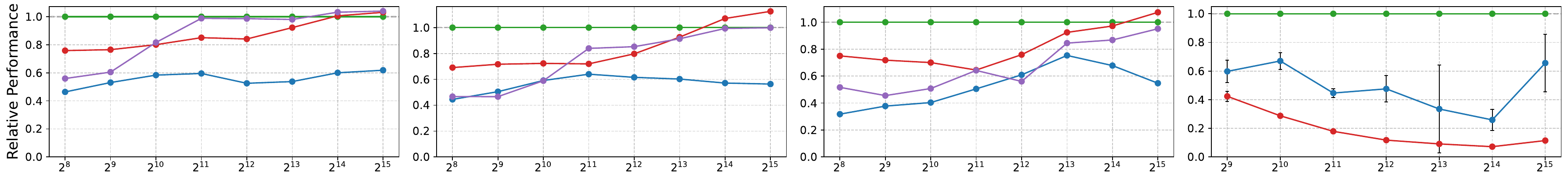}
    \vspace{-1.7em}
    \caption{GQA (DC) operator. Optimized with split-k update.}
  \end{subfigure}
  \begin{subfigure}[b]{0.95\textwidth}
    \includegraphics[width=\textwidth]{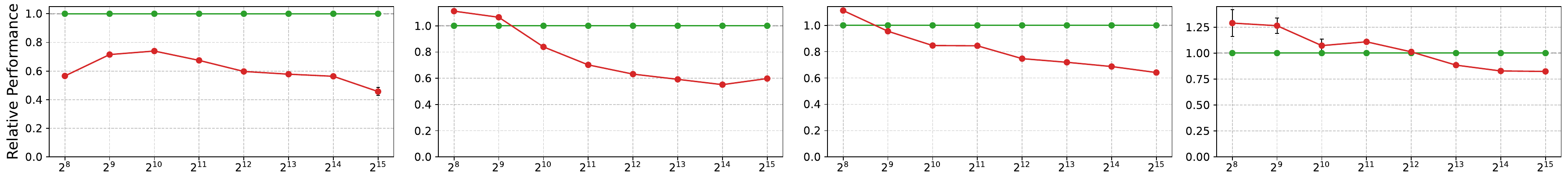}
    \vspace{-1.7em}
    \caption{ALiBi (PF) operator. Optimized with rolling update.}
  \end{subfigure}
  \begin{subfigure}[b]{0.95\textwidth}
    \includegraphics[width=\textwidth]{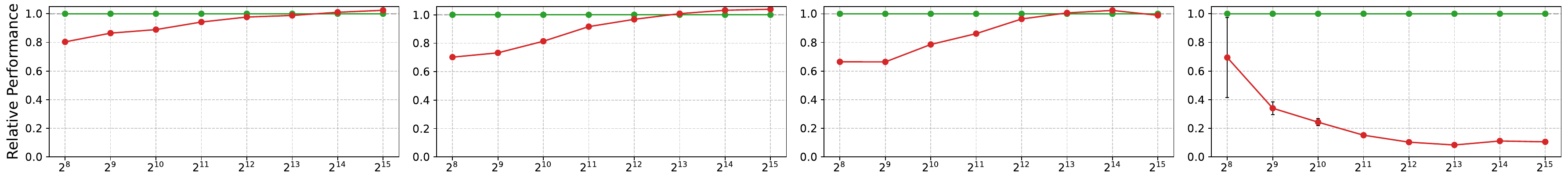}
    \vspace{-1.7em}
    \caption{ALiBi (DC) operator. Optimized with split-k update.}
  \end{subfigure}
  \begin{subfigure}[b]{0.95\textwidth}
    \includegraphics[width=\textwidth]{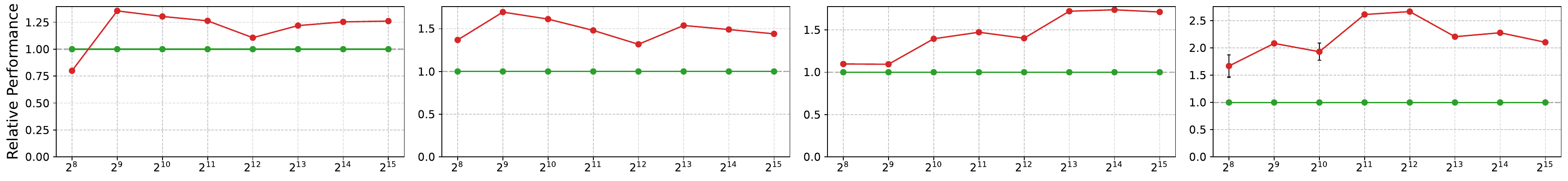}
    \vspace{-1.7em}
    \caption{Window (PF) operator. Optimized with rolling update.}
  \end{subfigure}
  \caption{Performance of \NAME{} vs.~tensor libraries on 8 operators for which library
  baselines exist (we were unable to find the library implementations of SoftCap PF/DC).}
  \label{fig:appendix-rq2-all-ops}
\end{figure*}

\clearpage

\subsection{Scalability Test}
Figure \ref{fig:appendix-throughput} extends the scalability test of Figure \ref{fig:throughput}
to all the 10 operators, 4 GPUs and all implementations used in our Evaluation section.
We still use a sequence length of 8192 and batch sizes up to 32.

Figures \ref{fig:appendix-throughput:pf-begin} to \ref{fig:appendix-throughput:pf-end}
shows that the throughput of all prefill kernels decrease as input batch size increases.
By closely inspecting profiling results, we find that
larger workloads are more likely to trigger GPUs' clock throttling.
Figure \ref{afig:nsys-screen} shows a profile in the Nvidia Nsight System profiler,
with a kernel (\texttt{flash\_fwd\_kernel}) that runs for $281$ milliseconds.
The GPU compute clock (``GPC Clock Frequency'') starts to throttle within $50$ milliseconds
after the kernel starts.

In Figures \ref{afig:thruput:dc-begin} to \ref{afig:thruput:dc-end},
the decoding kernels exhibit the opposite trend: throughput increases with batch size.
Decoding kernels are much less compute-bound than prefill kernels to trigger clock throttling,
and benefit from increasing parallelism at larger batch sizes.

\begin{figure*}[ht]
  \centering
  \begin{subfigure}[b]{0.95\textwidth}
    \includegraphics[width=\textwidth]{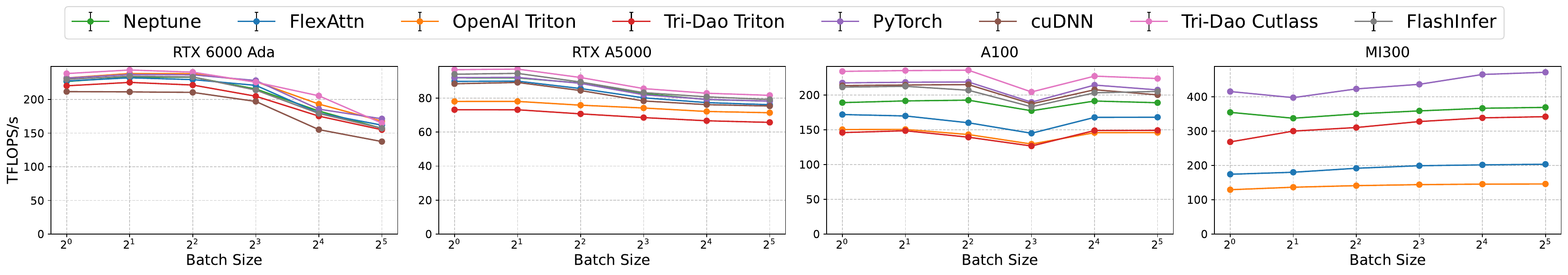}
    \vspace{-1.8em}
    \caption{Global (PF) operator. Optimized with rolling update.}
    \label{fig:appendix-throughput:pf-begin}
  \end{subfigure}
  \begin{subfigure}[b]{0.95\textwidth}
    \includegraphics[width=\textwidth]{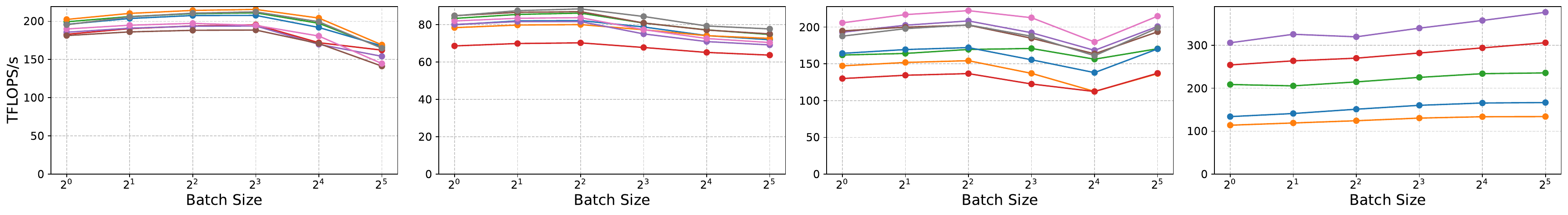}
    \vspace{-1.8em}
    \caption{Causal (PF) operator. Optimized with rolling update.}
  \end{subfigure}
  \begin{subfigure}[b]{0.95\textwidth}
    \includegraphics[width=\textwidth]{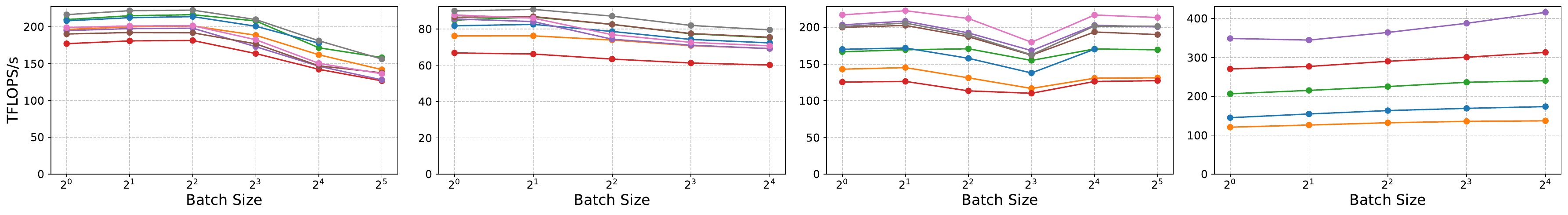}
    \vspace{-1.8em}
    \caption{GQA (PF) operator. Optimized with rolling update.}
  \end{subfigure}
  \begin{subfigure}[b]{0.95\textwidth}
    \includegraphics[width=\textwidth]{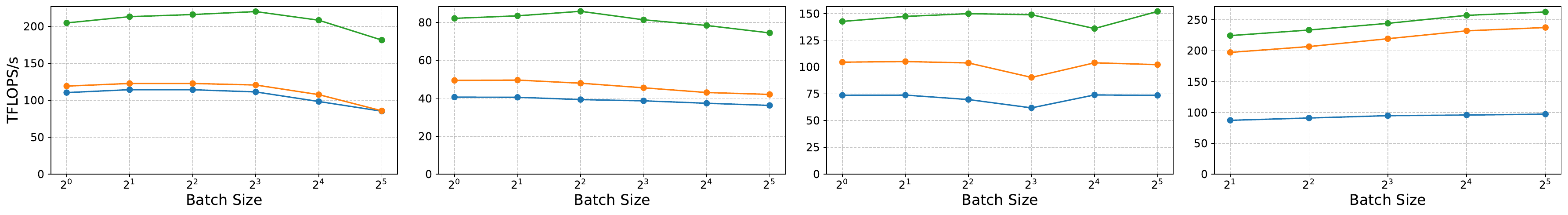}
    \vspace{-1.8em}
    \caption{ALiBi (PF) operator. Optimized with rolling update.}
  \end{subfigure}
  \begin{subfigure}[b]{0.95\textwidth}
    \includegraphics[width=\textwidth]{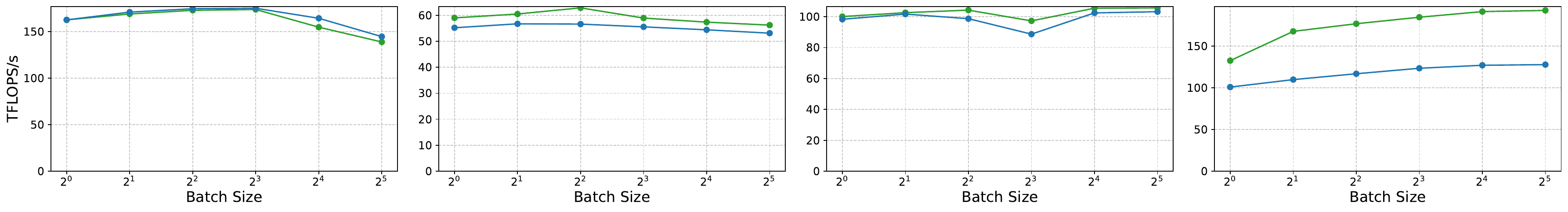}
    \vspace{-1.8em}
    \caption{SoftCap (PF) operator. Optimized with rolling update.}
  \end{subfigure}
  \begin{subfigure}[b]{0.95\textwidth}
    \includegraphics[width=\textwidth]{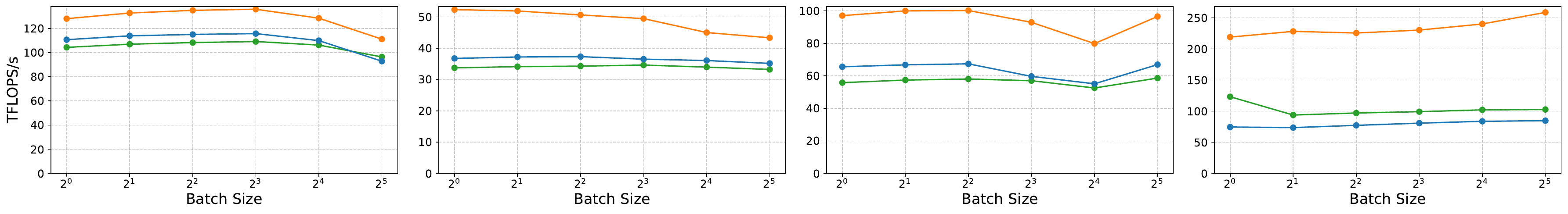}
    \vspace{-1.8em}
    \caption{Window (PF) operator. Optimized with rolling update.}
    \label{fig:appendix-throughput:pf-end}
  \end{subfigure}
  \caption{Throughput of \NAME{} kernels and multiple baselines over increasing batch size,
    for all 10 operators on all 4 GPUs.
  The y-axis shows throughput in TFLOPS/sec, and the x-axis shows batch size.}
  \label{fig:appendix-throughput}
\end{figure*}

\begin{figure}[t!]
  \ContinuedFloat
  \begin{subfigure}[b]{0.95\textwidth}
    \includegraphics[width=\textwidth]{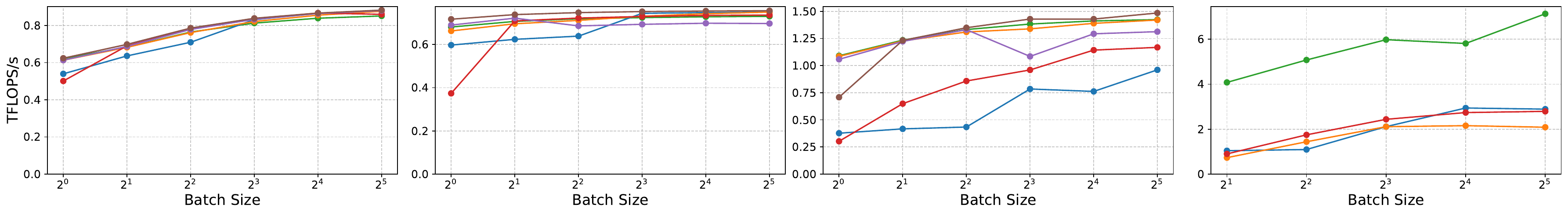}
    \vspace{-1.8em}
    \caption{Causal (DC) operator. Optimized with split-k update.}
    \label{afig:thruput:dc-begin}
  \end{subfigure}
  \begin{subfigure}[b]{0.95\textwidth}
    \includegraphics[width=\textwidth]{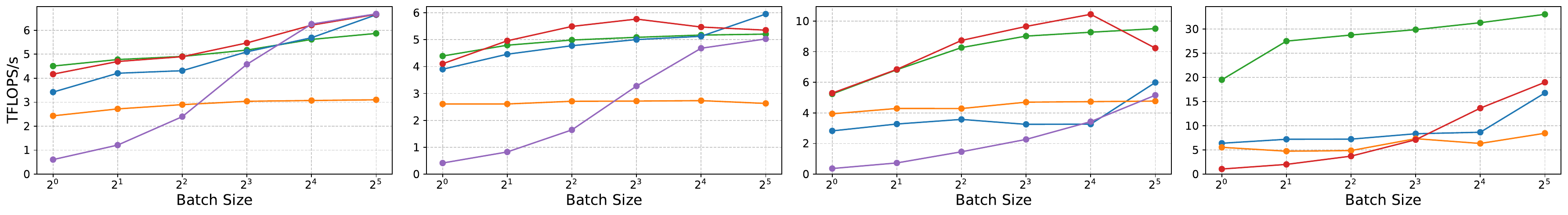}
    \vspace{-1.8em}
    \caption{GQA (DC) operator. Optimized with split-k update.}
  \end{subfigure}
  \begin{subfigure}[b]{0.95\textwidth}
    \includegraphics[width=\textwidth]{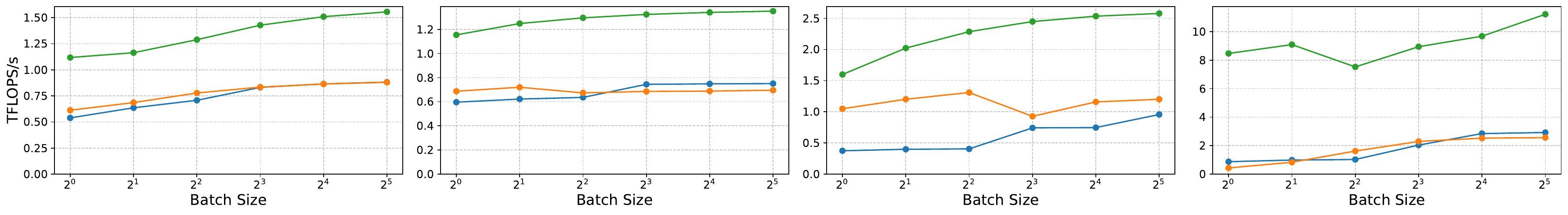}
    \vspace{-1.8em}
    \caption{ALiBi (DC) operator. Optimized with split-k update.}
  \end{subfigure}
  \begin{subfigure}[b]{0.95\textwidth}
    \includegraphics[width=\textwidth]{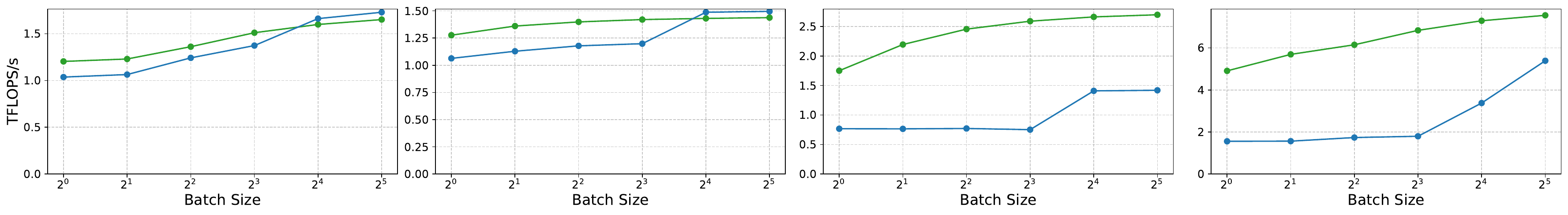}
    \vspace{-1.8em}
    \caption{SoftCap (DC) operator. Optimized with split-k update.}
    \label{afig:thruput:dc-end}
  \end{subfigure}
  \caption{(Continued) Throughput of \NAME{} kernels and multiple base-
    lines over increasing batch size,
    for all 10 operators on all 4 GPUs.
    The y-axis shows throughput in
  TFLOPS/sec, and the x-axis shows batch size.}
\end{figure}

\begin{figure}
  \centering\vspace{+.04in}
  \includegraphics[width=0.85\linewidth]{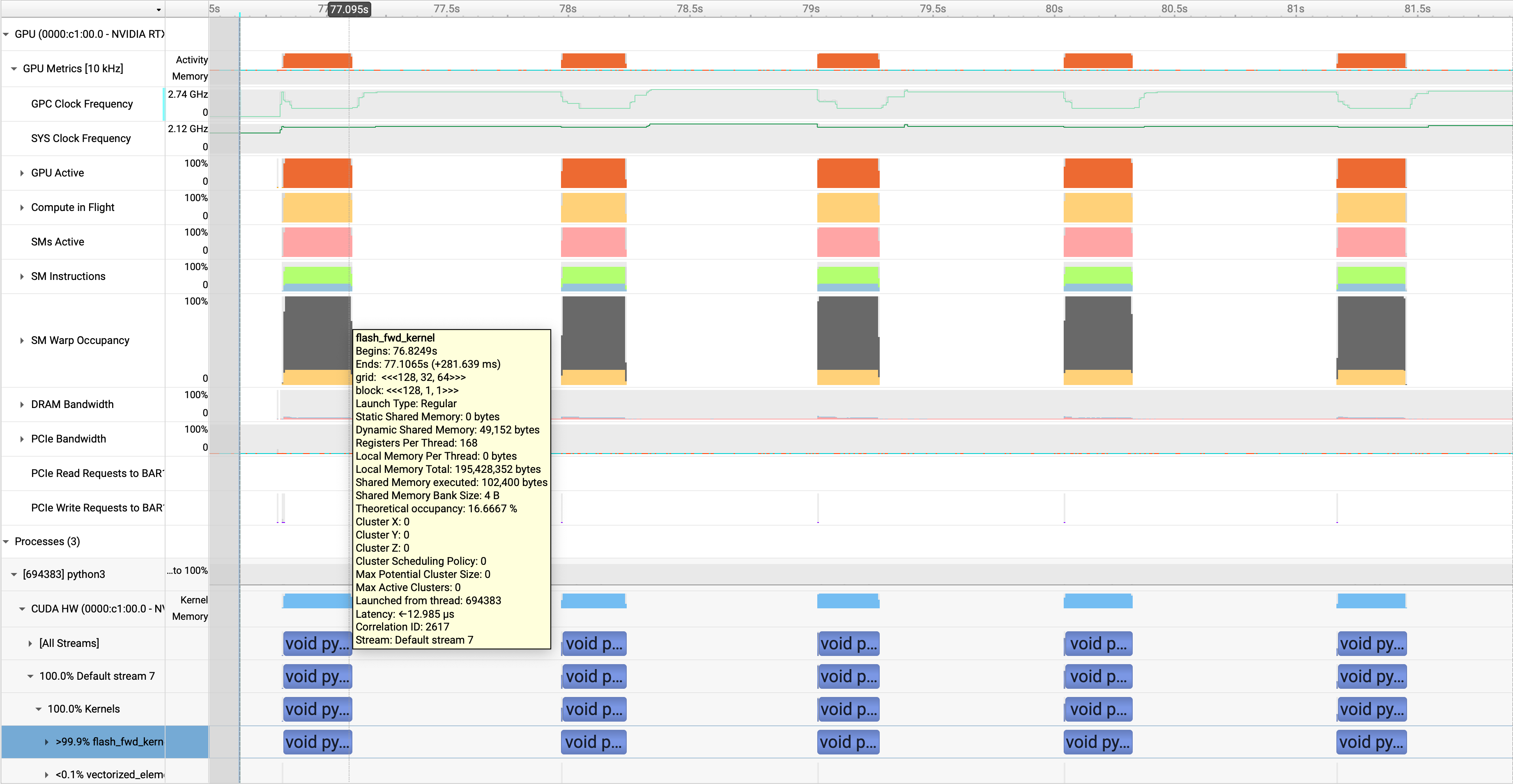}
  \caption{Timeline view of a profile in the Nvidia Nsight System profiler.
    The bottom rows show when the kernels are executing,
    and the ``GPC Clock Frequency'' row near the top shows the frequency of the GPU graphic clock
  over time.}
  \label{afig:nsys-screen}
\end{figure}

\clearpage

\section{Numerical Stability of \NAME{} Kernels}
\label{sec:app:num-stab}

\begin{table}[ht]
  \centering\small
  \caption{Numerical Error Summary (Absolute Error)}
  \label{tab:numeric_error_absolute_error}
  \begin{tabular}{lrrrrrrrrr}
    \hline
    Variant & \multicolumn{3}{c}{Neptune} & \multicolumn{3}{c}{Tri-Dao Attention} &
    \multicolumn{3}{c}{Flex Attention} \\
    & RMS & 90th \% & 99th \% & RMS & 90th \% & 99th \% & RMS & 90th \% & 99th \% \\
    \hline
    PF Global & 4.2e-05 & 1.5e-05 & 1.2e-04 & 4.2e-05 & 1.5e-05 & 1.2e-04 & 4.4e-05 &
    1.5e-05 & 1.2e-04 \\
    \hline
    PF Causal & 4.1e-05 & 1.9e-05 & 1.2e-04 & 4.2e-05 & 2.5e-05 & 1.2e-04 & 4.2e-05 &
    2.3e-05 & 1.2e-04 \\
    \hline
    PF GQA & 4.3e-05 & 2.3e-05 & 1.2e-04 & 4.5e-05 & 3.1e-05 & 1.2e-04 & 4.4e-05 &
    2.7e-05 & 1.2e-04 \\
    \hline
    PF ALiBi & 4.3e-05 & 1.5e-05 & 1.2e-04 & 5.4e-05 & 3.1e-05 & 2.4e-04 & 4.3e-05 &
    1.5e-05 & 1.2e-04 \\
    \hline
    PF Softcap & 2.4e-05 & 7.6e-06 & 3.1e-05 & -- & -- & -- & 2.4e-05 & 7.6e-06 & 3.1e-05 \\
    \hline
    PF Windowed & 2.4e-05 & 7.6e-06 & 3.1e-05 & 2.5e-05 & 7.6e-06 & 6.1e-05 & 2.4e-05 &
    7.6e-06 & 3.1e-05 \\
    \hline
    DC Causal & 3.9e-05 & 1.5e-05 & 1.2e-04 & 4.6e-05 & 1.5e-05 & 1.2e-04 & 3.0e-05 &
    1.5e-05 & 1.2e-04 \\
    \hline
    DC GQA & 3.4e-05 & 1.5e-05 & 1.2e-04 & 4.5e-05 & 1.5e-05 & 1.2e-04 & 2.6e-05 &
    7.6e-06 & 1.2e-04 \\
    \hline
    DC ALiBi & 4.3e-05 & 1.5e-05 & 1.2e-04 & 4.8e-05 & 1.5e-05 & 2.4e-04 & 4.1e-05 &
    1.5e-05 & 1.2e-04 \\
    \hline
    DC Softcap & 1.4e-05 & 1.9e-06 & 3.1e-05 & -- & -- & -- & 1.3e-05 & 9.5e-07 & 3.1e-05 \\
    \hline
  \end{tabular}
\end{table}

Optimizations that rewrite or re-associate floating-point operations may alter the
numerical properties of a kernel. We evaluate the numerical behavior of \NAME{}, Tri-Dao,
and FlexAttention kernels on a variety of attention operators, and find \NAME{} to
provide competitive numerical behavior. Using a set of common inputs, we evaluate kernel
outputs against an FP64 unfused attention kernel baseline.
Each kernel uses the same schedule and configuration as the performance evaluation.
We sample realistic attention inputs by running a pretrained Qwen2.5 model on 100
sentences with sequence length 2048 from
the WikiText dataset and recording the input tensors.

Table \ref{tab:numeric_error_absolute_error} presents the numerical error of all three frameworks,
represented by the root mean square error (RMS) and the magnitudes of the 90th and 99th
percentile errors. We also evaluate the distribution of per-sample (per-sentence) RMSE,
with results for GQA prefill and decoding show in Figure \ref{fig:gqa_rmse_combined}. For
all prefill benchmarks, Neptune kernels have lower error than the other two frameworks.
For the decode benchmarks, Neptune has lower
error than Tri-Dao Attention (when present) and comparable error to FlexAttention.

\begin{figure}[t]
  \centering
  \begin{subfigure}[t]{0.48\linewidth}
    \centering
    \includegraphics[width=\linewidth]{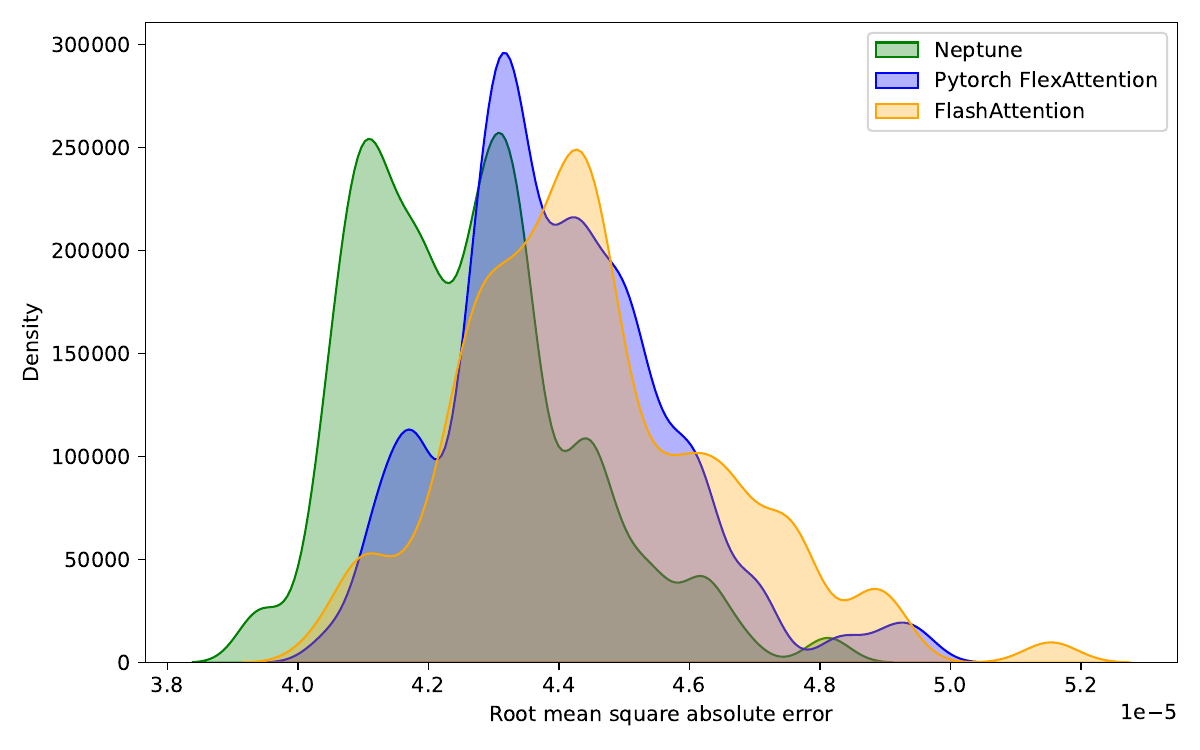}
    \caption{Prefill}
  \end{subfigure}
  \hfill
  \begin{subfigure}[t]{0.48\linewidth}
    \centering
    \includegraphics[width=\linewidth]{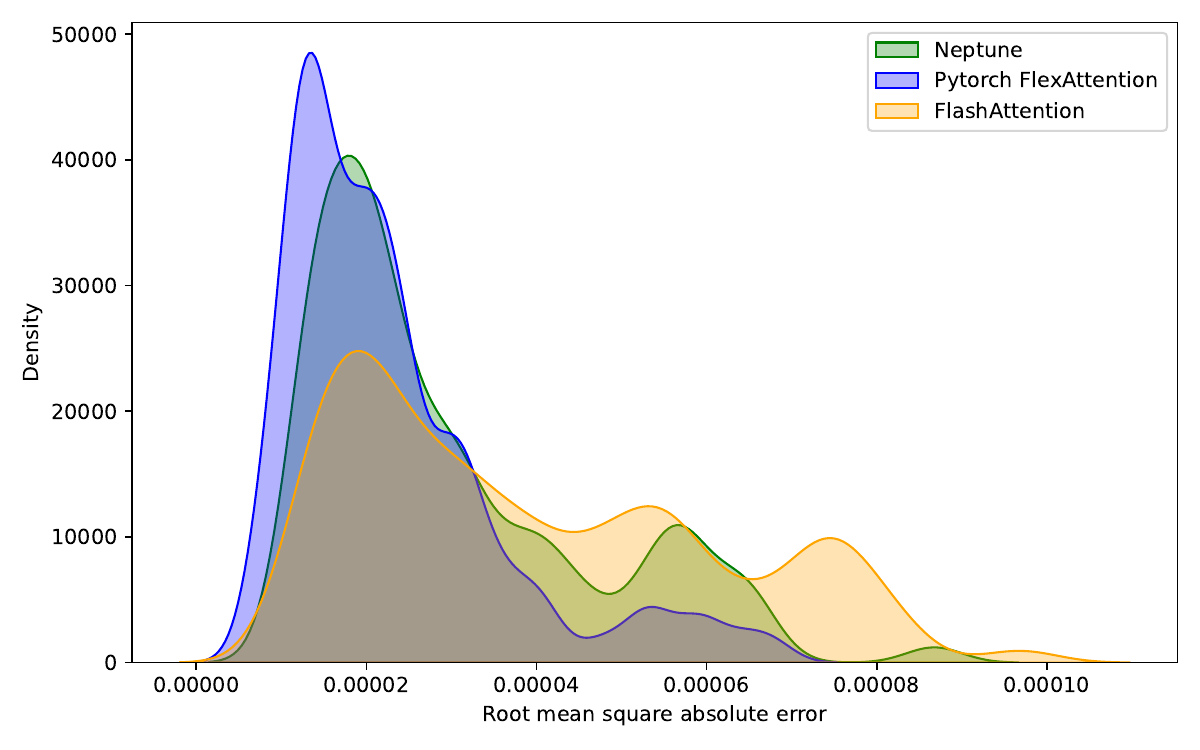}
    \caption{Decoding}
  \end{subfigure}
  \caption{Per-sample GQA RMSE for (a) prefill and (b) decoding phases.}
  \label{fig:gqa_rmse_combined}
\end{figure}
}

\end{document}